 \documentclass[aip,jcp,reprint,amsmath,amssymb,showpacs]{revtex4-1}%
\usepackage{hyperref}
\usepackage{graphicx}
 \usepackage{amsmath}
\usepackage{graphicx}%
\usepackage{amsfonts}%
 \usepackage{amssymb}
\usepackage{color}
\usepackage{setspace}
\usepackage{textcomp}

\begin{document}


\title{Scaling for rectification of bipolar nanopores as a function of a modified Dukhin number: the case of 1:1 electrolytes}

\author{D\'avid Fertig}
\author{Zs\'ofia Sarkadi}
\author{M\'onika Valisk\'{o}}
\author{Dezs\H{o} Boda}\email[Author for correspondence:]{boda@almos.vein.hu}
\affiliation{$^{1}$Center for Natural Sciences, University of Pannonia, P. O. Box 158, H-8201 Veszpr\'em, Hungary}
 
\begin{abstract}
The scaling behavior for the rectification of bipolar nanopores is studied using the Nernst-Planck equation coupled to the Local Equilibrium Monte Carlo method. The bipolar nanopore's wall carries $\sigma$ and $-\sigma$ surface charge densities in its two half regions axially. Scaling means that the device function (rectification) depends on the system parameters (pore length, $H$, pore radius, $R$, concentration, $c$, voltage, $U$, and surface charge density, $\sigma$) via a single scaling parameter that is a smooth analytical function of the system parameters. Here, we suggest using a modified Dukhin number, $\mathrm{mDu}=|\sigma|l_{\mathrm{B}}^{*}\lambda_{\mathrm{D}}HU/(RU_{0})$, where $l_{\mathrm{B}}^{*}=8\pi l_{\mathrm{B}}$, $l_{\mathrm{B}}$ is the Bjerrum length, $\lambda_{\mathrm{D}}$ is the Debye length, and $U_{0}$ is a reference voltage. We show how scaling depends on $H$, $U$, and $\sigma$ and through what mechanisms these parameters influence the pore's behavior. 
\end{abstract}

\maketitle


\section{Introduction}

\begin{figure}
\centering
\includegraphics[width=0.46\textwidth]{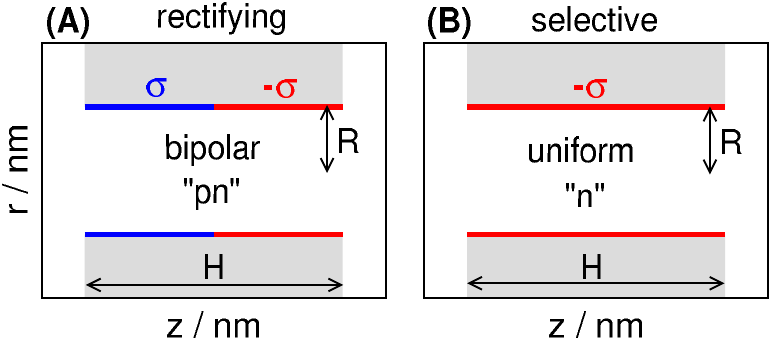}
\caption{The geometries of the nanopores under consideration. (A) The bipolar nanopore's device function is rectification. This is the system studied here. (B) The uniformly charged nanopore's device function is selectivity. The scaling behavior of this pore was studied in our previous work.~\cite{sarkadi_jcp_2021} The pores are cylindrical with radius $R=1$ nm and lengths $H=2-24$ nm.}
 \label{fig1-geom}
\end{figure}

Nanopores are channels in membranes connecting two baths that have radii comparable to the screening length, $\lambda$, of the electrolyte whose ions are transported through the pore. 
If the nanopore is asymmetric in the axial dimension along the pore, the ionic current is larger at one sign of the voltage (ON state) than at the opposite sign (OFF state), that is, the nanopore rectifies.
In this paper, we are interested in the case of electrostatically asymmetric cylindrical nanopores, specifically, bipolar nanopores \cite{daiguji_nl_2005,constantin_pre_2007,karnik_nl_2007,vlassiouk_nl_2007,kalman_am_2008,vlassiouk_acsnanno_2008,yan_nl_2009,cheng_acsnano_2009,nguyen_nt_2010,guo_acccr_2013,hato_pccp_2017,matejczyk_jcp_2017} whose wall carries positive surface charge density, $\sigma$, on one side and negative surface charge density, $-\sigma$, on the other side (Fig.~\ref{fig1-geom}A, $\sigma$ is always positive in this work).

In a previous publication~\cite{fertig_jpcc_2019}, we showed that, for fixed surface charge density, $\sigma$, pore length, $H$, and voltage, $U$, ionic current rectification defined as
\begin{equation}
 \mathrm{ICR}=\frac{I^{\mathrm{\mathrm{ON}}}}{I^{\mathrm{OFF}}}
 \label{eq:rect}
\end{equation} 
scales with the parameter $\xi = R/(\lambda z_{\mathrm{if}})$ where $R$ is the radius of the cylindrical nanopore, and $z_{\mathrm{if}}=\sqrt{z_{+}|z_{-}|}$ with $z_{+}$ and $z_{-}$ being the valences of the cations and the anions, respectively.
In Eq.~\ref{eq:rect}, $I^{\mathrm{ON}}$ and $I^{\mathrm{OFF}}$ are the magnitudes of the total currents at the forward and reverse biased voltages, respectively.
The device function of a device describes the relation between the output and input parameters of a device.
We prefer dimensionless device functions, such as ICR or selectivity, depending on the charge pattern of the nanopore.

By scaling, we mean that the device function (rectification, in this case) depends on the input variables via a single parameter that is a well-defined analytic function of the input variables:
\begin{equation}
 \mathrm{ICR}=\mathrm{ICR}\left[ \xi(R,c,z_{+},z_{-},\dots )\right] .
\end{equation} 
This scaling behavior may be useful not only for understanding the physics behind the device's behavior, but also for a practical purpose. If we know how ICR depends on $\xi$, for example, we can estimate the device function for a set of structural parameters that is difficult to test experimentally as nanopore fabrication is expensive. The predictive power of the phenomenon can be utilized in the design of nanodevices.

The $\xi$ parameter may depend on bulk concentration, $c$, via the screening length for which the Debye length 
\begin{equation}
\lambda_{\mathrm{D}} = 
\left( \dfrac{ c e^{2}}{\epsilon_{0}\epsilon kT} \sum_{i} z_{i}^{2}\nu_{i} \right)^{-1/2} ,
\label{eq:lambdaD}
\end{equation} 
is an obvious choice. In Eq.~\ref{eq:lambdaD}, $k$ is Boltzmann's constant, $T$ is the absolute temperature ($298.15$ K in this work), $e$ is the elementary charge, $\nu_{i}$ is the stoichiometric coefficient of ionic species $i$, $c$ is the electrolyte's bulk concentration that is related to the ionic bath concentrations via $c=c_{+}/|z_{-}|=c_{-}/z_{+}=c_{+}/\nu_{-}=c_{-}/\nu_{+}$, $\epsilon$ is the dielectric constant of the solvent ($78.45$ in this work), and $\epsilon_0$ is the permittivity of vacuum.
Another choice for the screening length is the one provided by the Mean Spherical Approximation (MSA) \cite{blum_mp_1975,blum_jcp_1977,nonner_bj_2000}, denoted by $\lambda_{\mathrm{MSA}}$.

The basic idea behind the scaling behavior is the following.
Double layers (DL) are formed at the charged wall of the nanopore and extend into the middle of the pore in the radial dimension.
They overlap in the centerline if $\lambda_{\mathrm{D}}/R\gg 1$. 
If they overlap, coions are excluded from the respective regions. 
The concept of coion is linked to the sign of the surface charge density, namely, in the positively charged region ($\sigma$, ``p'' region) the cation is the coion, while in the negatively charged region ($-\sigma$, ``n'' region) the anion is the coion.
Exclusion of an ionic species results in depletion zones for that ionic species.
A depletion zone means a low-concentration region along the pore.
It acts as a high-resistance element for the given ionic species along the pore axis if we imagine the consecutive segments of the pore as resistors connected in series.

In the short and narrow pores we consider, the mechanism of rectification is that the depletion zones of the coions are deeper in the OFF state of the voltage than in the ON state, namely, the concentrations are smaller in the OFF state than in the ON state.
Rectification scales with $\lambda_{\mathrm{D}}/R$ because $\lambda_{\mathrm{D}}/R$ tunes the degree of overlap of the DLs, and, consequently, the depth of the depletion zones.~\cite{fertig_jpcc_2019}
The OFF voltage can make an already deep depletion zone even more depleted and can produce rectification.
The importance of the $\lambda_{\mathrm{D}}/R$ ratio was also emphasized in numerous earlier studies.~ \cite{daiguji_nl_2005,vlassiouk_nl_2007,yan_nl_2009,albrecht_chapter_2013,Abgrall_2008,bocquet_csr_2010,daiguji_csr_2010,eijkel_csr_2010,zangle_csr_2010,madai_pccp_2018,dal_cengio_jcp_2019}

While $\sigma$ was constant in our study for the scaling of rectification~\cite{fertig_jpcc_2019}, its effect was examined separately in a study,~\cite{fertig_pccp_2020} where a non-monotonic $\sigma$-dependence of rectification was observed for electrolytes containing multivalent ions (2:2, 2:1, 3:1).
This behavior is caused by strong ionic correlations resulting in charge inversion, overcharging, and anion leakage.
These phenomena were reproduced by a hybrid simulation method that unifies the advantages of a simple transport equation and a particle simulation method, particularly, the Nernst-Planck (NP) equation and the Local Equilibrium Monte Carlo (LEMC) method that is a generalization of the Grand Canonical Monte Carlo (GCMC) method to non-equilibrium situations.
The methodology will be described in detail in section \ref{sec:method}.

The non-monotonic behavior was not reproduced by the Poisson-Nernst-Planck (PNP) theory that is a mean-field continuum theory using the Poisson-Boltzmann (PB) theory with the NP equation.
For 1:1 electrolytes, where ionic correlations are weak, the curves were monotonic and the NP+LEMC and PNP results agreed well.

In a recent paper,~\cite{sarkadi_jcp_2021} we considered a cylindrical nanopore with a uniform surface charge density on its wall (Fig.~\ref{fig1-geom}B).
For this nanopore, the obvious device function is selectivity that can be defined as $S_{+}=I_{+}/(I_{+}+I_{-})$, where $I_{i}$ is the ionic current carried by ionic species $i$.
We simulated nanopores of different radii, lengths, and surface charge density values, and also varied the concentration of the electrolyte and voltage.
We restricted ourselves to 1:1 electrolytes (as we do in this study) because we were curious when scaling works instead of when it does not work.

We found that identifying a ``universal'' scaling parameter that includes all the relevant parameters ($R$, $H$, $\sigma$, $c$, and $U$) is a real challenge.
A solid result that we reported is that the appropriate scaling parameter for the infinitely long ($H/R\rightarrow\infty$) pore (nanotube limit) is the Dukhin number defined as~\cite{bazant_pre_2004,chu_pre_2006,bocquet_chemsocrev_2010}
\begin{equation}
 \mathrm{Du}=\frac{|\sigma|}{eRc},
\end{equation} 
where $e$ is the elementary charge.
Du was originally introduced by Bikerman \cite{bikerman_1940} to characterize the ratio of the surface and volume conductances considering electrokinetic phenomena.
Later, Dukhin adopted the idea (see Ref.~\cite{dukhin_advcollsci_1993} and references therein) in studying electrophoretic phenomena.
His name was linked to the variable thanks to Lyklema who suggested the Du notation to salute Dukhin~\cite{lyklema_book_1995}, although the name `Bikerman number' is also present in the literature.
The Dukhin number proved to be useful in nanopore studies by replacing the colloid particle's radius with the pore radius to characterize a geometrical constraint in relation to an electrolyte property, the screening length. ~\cite{bazant_pre_2004,chu_pre_2006,khair_jfm_2008,das_langmuir_2010,bocquet_chemsocrev_2010,zangle_csr_2010,lee_nanolett_2012,yeh_ijc_2014,ma_acssens_2017,xiong_scc_2019,poggioli_jpcb_2019,dalcengio_jcp_2019,kavokine_annualrev_2020,noh_acsnano_2020}

We proposed using an alternative formula for Du that, for a 1:1 electrolyte, reads as
\begin{equation}
 \mathrm{Du}=\frac{|\sigma| l_{\mathrm{B}}^{*}\lambda_{\mathrm{D}}^{2}}{eR},
 \label{eq:du1}
\end{equation} 
where $l_{\mathrm{B}}^{*}=8\pi l_{\mathrm{B}}$ and $l_{\mathrm{B}}=e^{2}/4\pi \epsilon_0 \epsilon kT$ is the Bjerrum length.
It is based on the fact that the Debye length can be written as $\lambda_{\mathrm{D}}^{2}=1/(8\pi l_{\mathrm{B}} c)=1/(l_{\mathrm{B}}^{*}c)$. 
This form of Du makes it possible to relate quantities with the dimensions of distance to each other.
Also, we can use $\lambda_{\mathrm{MSA}}$ instead of $\lambda_{\mathrm{D}}$ in Eq.~\ref{eq:du1}  if it seems appropriate.
A screening length that depends on confinement~\cite{levy_jcis_2020} is also possible.

For the nanotube limit ($H/R\rightarrow\infty$), the variables $H$ and $U$ are absent; we used equilibrium calculations (PB and GCMC) to obtain selectivity.
For a finite nanopore, these variables become increasingly important because the DLs that appear at the entrances of the pore near the membrane on the two sides influence the ionic distributions inside the pore. 
In the nanohole limit ($H/R\rightarrow 0$), we found that a modified Dukhin number 
\begin{equation}
 \mathrm{mDu} = \mathrm{Du}\frac{H}{\lambda_{\mathrm{D}}} = \frac{|\sigma| l_{\mathrm{B}}^{*}\lambda_{\mathrm{D}}H}{eR}
 \label{eq:mDu}
\end{equation} 
is a much better scaling parameter.

Both the nanotube and nanohole limits are relevant experimentally.
The nanotube limit is obvious in cases, where a channel is fabricated in a thick membrane as in PET nanopores~\cite{Siwy_2004}.
The nanohole limit is relevant in the case of thin membranes, for example, graphene, MoS$_{2}$, or WS$_{2}$~\cite{garaj_n_2010,garaj_pnas_2013,OHern_nl_2014,rollings_nc_2016,thiruraman_nl_2018,thiruraman_acsnano_2020}.

If we plot $S_{+}$ as a function of $\mathrm{mDu}$ with a logarithmic scaling on the abscissa, e.g., $S_{+}$ vs.\ $\lg(\mathrm{mDu})$, the resulting curve is a sigmoid whose inflection point has a basic importance.
First, the inflection point separates state points for which surface conduction dominates from state points where volume (bulk) conduction dominates.
$S_{+}\sim 0.5$ corresponds to the non-selective cases, when volume conduction dominates; both cations and anions contribute to the current.
$S_{+}\sim 1$ corresponds to the selective cases, when surface conduction dominates; the DL extends over the pore in the $r$ dimension and only the counterions contribute to the current.
The inflection point (where $S_{+}\approx 0.75$) offers itself to be a transition point between these limiting cases.

When we talk about scaling, however, we are interested in the cases in between where $S_{+}$ is intermediate.
The inflection point is a useful asset because if the inflection points of various scaling curves are at the same $\lg(\mathrm{mDu})$ values, we can say that the curves fall onto each other.
Such a role of the inflection point was not found for bipolar nanopores.

The limiting behavior of the voltage dependence for $U\rightarrow 0$ is different in selective and bipolar nanopores.
In the selective case (uniform $\sigma$, Fig.~\ref{fig1-geom}B), the $U\rightarrow 0$ limit provides a limiting value for selectivity,
\begin{equation}
\lim_{U\rightarrow 0} S_{+}(U) = S_{+}^{0},
\end{equation} 
the slope conductance limit.
In Ref.~\cite{sarkadi_jcp_2021} we reported a quadratic dependence of the inflection point on voltage. 
Voltage, however, is absent in Eq.~\ref{eq:mDu}.
Including $U$ in the scaling parameter requires more data.

In the rectifying case (bipolar $\sigma$, Fig.~\ref{fig1-geom}A), rectification is based on the fact that the voltage has different effects in the ON and OFF cases.
Larger voltage produces larger rectification because larger voltage produces larger differences between the concentration profiles in the ON and OFF cases.
Accordingly, rectification vanishes in the $U\rightarrow 0$ limit:
\begin{equation}
 \lim_{U\rightarrow 0} \mathrm{ICR}(U) = 1 .
\end{equation} 
We will show in this paper that $\sigma$ and $U$ have similar roles in the bipolar nanopore: they both tune rectification.
They do it with different mechanisms, but from the point of view of scaling, with similar results.
Therefore, we redefine mDu for bipolar nanopores as
\begin{equation}
 \mathrm{mDu} =  \frac{|\sigma| l_{\mathrm{B}}^{*}\lambda_{\mathrm{D}}H}{eR} \frac{U}{U_{0}} ,
 \label{eq:mDu_new}
\end{equation} 
where $U_{0}$ is an arbitrary voltage to make mDu dimensionless ($U_{0}=200$ mV in this work).
For $U=200$ mV (a value that we used in most of our simulations), Eqs.~\ref{eq:mDu} and \ref{eq:mDu_new} provide the same value. 

The length of the nanopore, $H$, also influences the device behavior differently in the selective and bipolar cases.
Because bipolar nanopores are necessarily finite due to their asymmetry, the nanotube limit ($H\rightarrow\infty$) does not make sense in this case.
Indeed, the original Dukhin number is not an appropriate scaling parameter for the bipolar nanopore.
In this paper, we examine whether mDu is better for this purpose.

The length of the pore influences the selective pore's behavior in the middle of the pore because the DLs formed at the entrances extend in the pore in the axial dimension; the degree of this extension is characterized by $\lambda_{\mathrm{D}} /H$.
The behavior of the bipolar nanopore is also sensitive to $H$. 
The depletion zones are formed at the junction of the two differently charged regions (``n'' and ``p'') and they are deeper in the OFF state if the nanopore is longer.

In this paper, we examine the applicability of a scaling parameter ($\mathrm{mDu}$, Eq.~\ref{eq:mDu_new}) that includes all relevant system parameters, $R$, $H$ $c$, $\sigma$, and $U$.
Even if scaling is not perfect, showing our results in terms of such a parameter is a compressed way of analyzing the results.
This is useful even if we express the device function in a subset of the parameter space, e.g., changing only a few (one or two) of all the parameters, while keeping the rest of them fixed.

The device function is rectification (Eq.~\ref{eq:rect}) in this study.
It proved to be useful in our previous work~\cite{sarkadi_jcp_2021} that selectivity was defined as a bounded function ($0.5<S_{+}<1$) that produced a sigmoid curve as a function of $\lg(\mathrm{mDu})$.
Following the practice of that paper, we pursue sigmoid curves in this work as well, and introduce the following definition for rectification:
\begin{equation}
 \mathrm{ICR}' = \frac{\mathrm{ICR}-1}{\mathrm{ICR}+1} = \frac{I^{\mathrm{ON}}-I^{\mathrm{OFF}}}{I^{\mathrm{ON}}+I^{\mathrm{OFF}}}.
\end{equation} 
This function is $1$ for perfect rectification ($\mathrm{ICR}\rightarrow \infty$) and $0$ for no rectification ($\mathrm{ICR}=1$).

\section{Model and method}
\label{sec:method}

The model of the nanopore/membrane/baths system has been constructed of two baths separated by a membrane and connected with a cylindrical pore through the membrane.
The results presented in this paper will be expressed as functions of cylindrical coordinates ($z$ and $r$) due to the rotational symmetry of the system around the axis of the pore.
Throughout the paper, if $z$ has a subscript, it denotes ionic valence, while without a subscript it denotes the axial coordinate.
The simulation domain is a cylinder with a $15-30$ nm length and $8-9$ nm radius (depending on $c$ and $H$), while the pore's length, $H$, is varied between $2$ and $24$ nm.
The membrane and the pore are confined by hard walls with a surface charge placed on the wall of the pore (Fig.~\ref{fig1-geom}). 
The radius of the pore is constant in this study ($R=1$ nm).

The pore has two regions: a positively charged region with surface charge density $\sigma$ and a negatively charged region with surface charge density $-\sigma$; the length of each is $H/2$.
The surface charge density on the cylinder is modeled as a collection of fractional point charges on a $0.2\times 0.2$ nm grid ($0.3 \times 0.3$ for longer pores above $H=15$ nm.)

The electrolyte is modeled in the implicit solvent framework.
This means that the effect of the water molecules on the ions is taken into account implicitly by response functions.
One response function is the dielectric response (screening) represented by the dielectric constant, $\epsilon$, of the electrolyte.
In our study, it is constant system-wide and appears in the denominator of the Coulomb potential.
The interaction potential between ions can be expressed as the sum of the hard-sphere potential and the Coulomb potential:
\begin{equation}
 u_{ij}(r)=
 \left\lbrace 
\begin{array}{ll}
\infty & \quad \mathrm{if} \quad r<(d_{i}+d_{j})/2 \\
 \dfrac{1}{4\pi\epsilon_{0}\epsilon} \dfrac{z_{i}z_{j}e^{2}}{r} & \quad \mathrm{if} \quad r \geq (d_{i}+d_{j})/2\\
\end{array}
\right. 
\label{eq:uij}
\end{equation}
where $d_i$ and $d_j$ are the ionic diameters ($d_{+}=d_{-}=0.3$ nm in this study), $z_i$ and $z_j$ are the valences of the different ionic species, and $r$ is the distance between ions $i$ and $j$. 

The other response function is dynamic in nature; the water molecules collide with ions and impede their diffusion. 
We take that effect into account by a diffusion coefficient function, $D_{i}(\mathbf{r})$, in the NP equation that reads as
\begin{equation}
 \mathbf{j}_{i}(\mathbf{r})= -\frac{1}{kT} D_{i}(\mathbf{r})c_{i}(\mathbf{r})\nabla \mu_{i}(\mathbf{r}),
 \label{eq:np}
\end{equation} 
where $\mathbf{j}_{i}(\mathbf{r})$, $D_{i}(\mathbf{r})$, $c_{i}(\mathbf{r})$, and $\mu_{i}(\mathbf{r})$ are the flux density, the diffusion coefficient profile, the concentration profile, and the electrochemical potential profile of ionic species $i$, respectively.
The concentration profile, $c_{i}(\mathbf{r})$, must be distinguished from the bulk concentration, $c_{i}$.
We always indicate the function's argument as $c_{i}(\mathbf{r})$ or $c_{i}(z)$ (if it is cross-section averaged) if we talk about a profile.

For the diffusion coefficient profile, $D_{i}(\mathbf{r})$, we use a piecewise constant function, where the value in the baths is $1.334\times 10^{-9}$ m$^{2}$s$^{-1}$ for both ionic species, while it is the tenth of that inside the pore, $D_{i}^{\mathrm{pore}}$, as in our earlier works.~\cite{matejczyk_jcp_2017,madai_jcp_2017,madai_pccp_2018,fertig_jpcc_2019,fertig_pccp_2020,sarkadi_jcp_2021}
These particular choices do not qualitatively affect our conclusions.

The electrochemical potential profile can be expressed as 
\begin{equation}
 \mu_{i}(\mathbf{r}) = \mu_{i}^{0} + kT\ln c_{i}(\mathbf{r}) + \mu_{i}^{\mathrm{BMF}} (\mathbf{r}) + z_{i}e\Phi(\mathbf{r}),
 \label{eq:elchempot}
\end{equation} 
where  $\mu_{i}^{0}$ is a temperature dependent reference term (it carries the information about the unit of $c_{i}(\mathbf{r})$; strictly speaking, $\ln c_{i}$ should be written as $\ln (c_{i}/c_{0})$, where $c_{0}$ is a reference concentration also contained by $\mu_{i}^{0}$), $\Phi(\mathbf{r})$ is the mean (ensemble or time averaged) electrical potential, and $\mu_{i}^{\mathrm{BMF}} (\mathbf{r})$ is an excess term beyond the mean field (BMF) approximation. 

The chemical and electrical contributions, $\mu_{i}^{\mathrm{CHEM}}(\mathbf{r})$ and $\mu_{i}^{\mathrm{EL}}(\mathbf{r})$, can be defined formally even if they are difficult to separate in experiments.
The chemical term is 
\begin{equation}
 \mu_{i}^{\mathrm{CHEM}}(\mathbf{r}) = \mu_{i}^{0} + kT\ln c_{i}(\mathbf{r}) + \mu_{i}^{\mathrm{BMF}} (\mathbf{r}),
\label{eq:chempot}
 \end{equation} 
while the electrical term is 
\begin{equation}
 \mu_{i}^{\mathrm{EL}}(\mathbf{r}) = z_{i}e\Phi(\mathbf{r}).
\label{eq:elterm}
 \end{equation} 
The BMF and EL terms together constitute the excess term.
 
The ion transport is steady-state; the profiles in the above equation do not depend on time.
The transport is maintained by constant (time-independent) boundary conditions for the concentration and the electrical potential on the two sides of the membrane on the two half cylinders confining the simulation cell from left and right.

Fixing the concentration ($c_{i}^{\mathrm{L}}$ and $c_{i}^{\mathrm{R}}$) and the electrical potential ($\Phi^{\mathrm{L}}$ and $\Phi^{\mathrm{R}}$) on the left (L) and right (R) boundaries means fixing the electrochemical potentials that can be written as
\begin{equation}
 \mu_{i}^{\mathrm{L}}=\mu_{i}^{0} + kT\ln c_{i}^{\mathrm{L}} + \mu_{i}^{\mathrm{BMF, L}} + z_{i}e\Phi^{\mathrm{L}}
 \label{eq:left}
\end{equation}
and
\begin{equation}
 \mu_{i}^{\mathrm{R}}=\mu_{i}^{0} + kT \ln c_{i}^{\mathrm{R}} + \mu_{i}^{\mathrm{BMF, R}} + z_{i}e\Phi^{\mathrm{R}},
 \label{eq:right}
\end{equation}
where $\mu_{i}^{\mathrm{BMF, L}}$ and $\mu_{i}^{\mathrm{BMF, R}}$ are the excess chemical potentials in the absence of an external field determined by the Adaptive GCMC method of Malasics et al.  \cite{malasics_jcp_2010}.
The  $z_{i}e\Phi^{\mathrm{L}}$ and $z_{i}e\Phi^{\mathrm{R}}$ terms are the interactions with the applied electrical potentials in the two baths.
Prescribing $\Phi^{\mathrm{L}}$ and $\Phi^{\mathrm{R}}$ on the system's boundary means that we use an electrostatic Dirichlet boundary condition.
Voltage is defined as $U=\Phi^{\mathrm{R}}-\Phi^{\mathrm{L}}$ in our study.

To make use of the NP equation, we need a relation between the concentration profile, $c_{i}(\mathbf{r})$, and the electrochemical potential profile, $\mu_{i}(\mathbf{r})$.
Most of our results reported in this work have been obtained by the LEMC technique developed by Boda and Gillespie~\cite{boda_jctc_2012}. 
It is a variant of the GCMC method that is generally applied in global equilibrium ($\mu_{i}$ is constant in space), while only local equilibrium is assumed in LEMC in subvolumes of the simulation cell.
Inside these subvolumes, $\mu_{i}(\mathbf{r})$ is constant, but it varies between subvolumes.
The resulting varying $\mu_{i}(\mathbf{r})$ profile provides a driving force for the ionic transport.
Coupled with NP (denoted by NP+LEMC), an iterative procedure is obtained that provides a solution where $c_{i}(\mathbf{r})$ is statistical-mechanically consistent with $\mu_{i}(\mathbf{r})$ and also they together produce a flux via the NP equation that satisfies conservation of mass ($\nabla \cdot \mathbf{j}(\mathbf{r})=0$). 

The LEMC simulations use the interaction potentials given in Eq.~\ref{eq:uij} also called charged hard spheres or the primitive model of electrolytes. 
Overlap of hard-sphere ions with each other and with the hard walls in the system (wall of the pore, the membrane, and the system's outer boundary) is forbidden.

The other method, PNP, uses the PB theory to relate $c_{i}(\mathbf{r})$ to $\mu_{i}(\mathbf{r})$.
In this case, explicit particles are not present in the calculation; ions are described by the $c_{i}(\mathbf{r})$ functions.
The hard-sphere cores are absent, so the ions are treated as point charges.
When we couple the PB theory to the NP equation, the PNP theory is obtained that is a mean-field continuum theory applied extensively to bipolar nanopores  \cite{daiguji_nl_2005,constantin_pre_2007,vlassiouk_nl_2007,karnik_nl_2007,vlassiouk_acsnanno_2008,kalman_am_2008,yan_nl_2009,nguyen_nt_2010,szymczyk_jpcb_2010,singh_jap_2011,singh_jpcb_2011,singh_apl_2011,oeffelen_oneplos_2015,tajparast_bba_2015}.
Note that the NP equation has also been coupled to a density functional theory by Gillespie.~\cite{gillespie_jpcm_2002,gillespie_pre_2003,gillespie_jpcb_2005,gillespie_bj_2008_energetics}

Being mean field means that the ions interact with other ions only via the mean electrical potential produced by them.
Formally, this means that the BMF term in Eq.~\ref{eq:elchempot} is zero.
The resulting electrochemical potential is that of an ideal solution.
The PNP theory is quite standard in nanopore studies and is described elsewhere in detail.~\cite{constantin_pre_2007,burger_nonlin_2012,eisenberg_pnp_2019}
Our procedure has been detailed in Ref.~\cite{matejczyk_jcp_2017}.

Because the NP+LEMC method is less standard, we provide a brief description in Appendix \ref{sec:appendix}, while more details can be found in our earlier papers.~\cite{boda_jctc_2012,boda_arcc_2014,boda_jml_2014,fertig_hjic_2017}

\section{Results and Discussion}
\label{sec:res}

\subsection{Mechanism of rectification}

\begin{figure}
\centering
\includegraphics[width=0.35\textwidth]{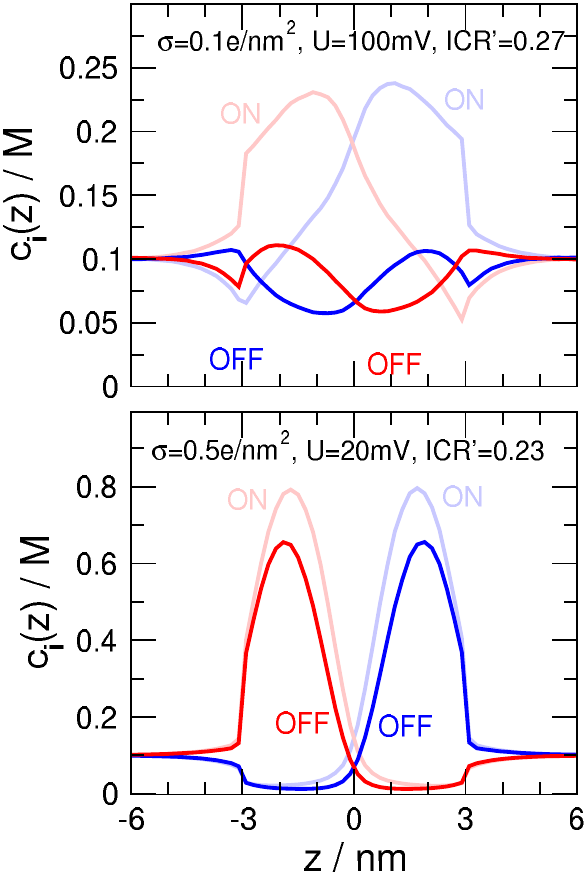}
\caption{Axial concentration profiles for a fixed concentration, $c=0.1$ M, pore radius, $R=1$ nm, and pore length, $H=6$ nm. 
The surface charge density, $\sigma$, and voltage, $U$, are chosen in the two panels in such a way that their product is constant; $\sigma=0.1$ $e$/nm$^{2}$ and $U=100$ mV in the top panel, while $\sigma=0.5$ $e$/nm$^{2}$ and $U=20$ mV in the bottom panel ($\mathrm{mDu}=5.16$ in both cases, Eq.~\ref{eq:mDu_new}).
Blue and red lines refer to cation and anion profiles, respectively.
Red and light red (blue and light blue) profiles refer to OFF and ON states, respectively.
In general, blue and red colors refer to cations and anions, respectively, here and later figures. 
} 
\label{fig2:c}
\end{figure}

Let us introduce our discussion with axial concentration profiles to pin down the mechanism by which a bipolar nanopore rectifies ionic current.
The cross-section-averaged concentration profile is defined as
\begin{equation}
c_{i}(z) =  \dfrac{1}{(R_{\mathrm{lim}}(z)-d_{i}/2)^{2}\pi}  \int\limits_{0}^{R_{\mathrm{lim}}(z)-d_{i}/2} c_{i}(z,r) 2\pi r \mathrm{d} r ,
\label{eq:cz}
\end{equation}
where $R_{\mathrm{lim}}(z)=R$ for $-H/2<z<H/2$, while it is the radius of the simulation cell outside the pore ($|z|>H/2$). 
Fig.~\ref{fig2:c} shows that changing the sign of the voltage from ON to OFF results in reduced concentrations inside the pore. 
Regions where concentrations are reduced, are generally termed depletion zones in the semiconductor, nanopore, and ion channel literature.
Depletion zones of coions are formed as results of the surface charges that repulse the coions. 
In the ``p'' zone cations have depletion zones (they are the coions), while in the ``n'' zone the anions have depletion zones (they are the coions).
The basis of the mechanism of rectification is that the OFF voltage makes the depletion zones deeper compared to the ON state.

This indicates that in the following discussions we need to pay extra attention to the depletion zones in the OFF state because tuning certain parameters ($\sigma$, $c$, or $H$, for example) changes rectification via modifying the depletion zones.
We need to state right at the beginning, however, that influencing the ON-state profiles is also a legal way to tune rectification.
Indeed, as we will show, increasing voltage have a stronger effect on increasing the ON-state concentration profiles than the OFF-state concentration profiles.

Fig.~\ref{fig2:c} shows that the $c_{i}(z)$ profiles behave differently in the top and the bottom panels, still, rectification is similar in the two cases.
The two panels refer to the same $\mathrm{mDu}$ (same $\sigma U$ product), but different pairs of $\sigma$ and $U$.
These two pairs of $\sigma$ are $U$ were chosen so that the rectification is neither too large, nor too small.
The agreement of rectifications indicates that mDu may be an appropriate scaling parameter, but we need to examine the scaling curves to show this.

\subsection{Scaling curves}

Next, let us turn our attention to the device-level behavior, namely, let us show the device function (rectification) as a function of the chosen scaling parameter, mDu.
As it was stated before, the scaling parameter does not need to be a perfect one. 
Our main premise here is that even an imperfect scaling parameter is useful in explaining the behavior of the device and in analyzing the effects of the various parameters. 

\begin{figure*}
\centering
\includegraphics[width=0.6\textwidth]{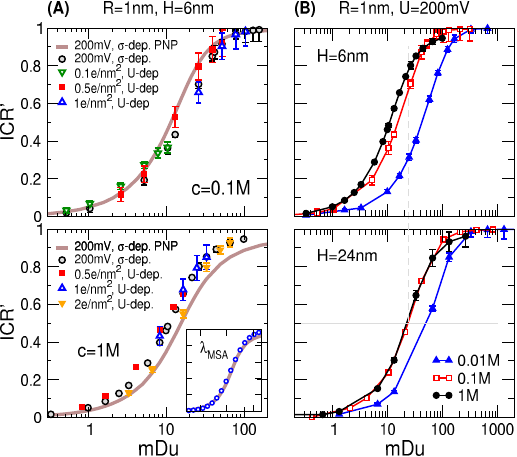}
\caption{Rectification, $\mathrm{ICR}'$, as a function of $\mathrm{mDu}=|\sigma|l_{\mathrm{B}}^{*}\lambda_{\mathrm{D}}HU/(RU_{0})$, where $U_{0}=200$ mV. In panel (A) pore radius and pore length are kept fixed at $R=1$ nm and $H=6$ nm, while concentrations are $c=0.1$ M (top panel) and $c=1$ M (bottom panel).
The various symbols/curves have been obtained by either changing $\sigma$ in the range of $|\sigma|=0.001-2$ $e$/nm$^{2}$ at fixed $U$ (see curves denoted by $\sigma$-dep.) or by changing $U$ in the range of $|U|=10-200$ mV at fixed $\sigma$ (see curves denoted by $U$-dep.).
The inset in the bottom panel for $c=1$ M was obtained by using $\lambda_{\mathrm{MSA}}=0.415$ nm instead of $\lambda_{\mathrm{D}}=0.304$ nm in the case of the NP+LEMC data.
The open blue symbols were obtained by merging all the NP+LEMC data and fitting a sigmoid. 
In panel (B) pore radius and voltage are kept fixed at $R=1$ nm and $U=200$ mV, while pore lengths are $H=6$ nm (top panel) and $H=24$ nm (bottom panel).
The various symbols/curves refer to different concentrations and have been obtained by changing $\sigma$  (in the range of $|\sigma|=0.001-3$ $e$/nm$^{2}$).
Error bars have been obtained from the variances of the total current values over the iterations for which we averaged and from the error propagation law for ICR'.}
\label{fig3:scaling}
\end{figure*}

Since we have many system parameters ($R$, $H$, $c$, $U$, and $\sigma$ and there would be even more for multivalent electrolytes such as $z_{+}$ and $z_{-}$),  it seems to be an  insurmountable task to find the Holy Grail of scaling parameters, a parameter as a function of which all the rectification curves fall onto each other perfectly.
Du was found to be such a ``perfect'' scaling parameter for the infinitely long selective pore in our previous work,~\cite{sarkadi_jcp_2021} but deviations were found when we decreased the length of the pore.
An interplay between radial and axial effects took place that also occurs in the bipolar pore considered here.

So our purpose here is to show that changing $\sigma$, $U$, and $H$ have similar effects on the behavior of the nanopore.
They are all in the numerator of mDu, so if scaling is valid, doubling any of these parameters will double rectification. 
Although such an accurate scaling is not present, we can put forward a weaker statement that increasing any of these parameters will increase rectification.

Concentration, for a given electrolyte, determines the screening length, so it determines the degree of overlap of the DLs formed at the nanopore's wall.
That way, the ratio of the screening length and pore radius, $\lambda_{\mathrm{D}} /R$ , determines rectification if everything else is constant as it was shown in our earlier publication.~\cite{fertig_jpcc_2019}
Because this question was analyzed in detail in that paper, we pay less attention to it here;  instead, we fix the pore radius at $R=1$ nm and study the device behavior at various fixed concentrations, $c=0.01$, $0.1$, and $1$ M, with $\sigma$, $H$, and $U$ being the main variables.

Fig.~\ref{fig3:scaling} shows scaling curves under different circumstances.
Fig.~\ref{fig3:scaling}A shows that we obtain similar behavior either by changing $\sigma$ or $U$.
As we will show through the behavior of the ionic distributions later in this work, increasing any of these two parameters increases rectification.
This figure, furthermore, implies linearity: it does not matter if we change $\sigma$ or $U$ to the same degree, we get the same effect although via different mechanisms as implied by Fig.~\ref{fig2:c}.
Fig.~\ref{fig3:scaling}A suggests that mDu is a fair scaling parameter as far as the relation of $\sigma$ and $U$ is concerned.
A close examination of the differences between the two panels, however, reveals that we have different scaling behaviors for $c=0.1$ M and $1$ M.

This difference is better shown in Fig.~\ref{fig3:scaling}B that shows scaling curves for different concentrations inside one panel for a fixed voltage, $U=200$ mV.
Indeed, the different curves for $c=0.01$, $0.1$, and $1$ M in the top panel for $H=6$ nm are relatively far apart.
The $c=0.1$ and $1$ M curves are closer, while the $c=0.01$ M curve deviates more.
This is the result of the fact that the screening lengths at the studied concentrations are comparable in size to the length and radius of the pore; the Debye lengths are $\lambda_{\mathrm{D}}=3.042$, $0.962$ , and $0.304$ nm for $c=0.01$, $0.1$, and $1$ M, respectively.
Such wide DLs couple the axial and radial behaviors of the ionic distributions.

If the insight about the effect of the axial interaction between the ``p'' and ``n'' regions in short pores is right, we should observe better scaling in longer pores.
Indeed, the bottom panel of Fig.~\ref{fig3:scaling}B for $H=24$ nm shows that the curves for $c=0.1$ and $1$ M scale appropriately. 
The pore is long enough compared to the screening lengths of the $c=0.1$ and $1$ M electrolytes.
The $c=0.01$ M curve is still off, but to a lesser extent than in the shorter pore.

The main message of Fig.~\ref{fig3:scaling}B is that linear scaling as a function of $H$ ($\mathrm{mDu}\sim H$) works quite well, though it is far from being perfect. 
On the basis of the available results, we may predict that scaling would improve with increasing $H$, but at this point of the research, this is only a hypothesis.
A further complication of the picture is that $\lambda_{\mathrm{D}}$ is not necessarily a good screening length in confinement as implied by the study of Levy et al.\ \cite{levy_jcis_2020}.

For simplicity, we use the Debye length for $\lambda$ in this study for both NP+LEMC and PNP.
In the inset of the bottom panel of Fig.~\ref{fig3:scaling}A, however, we also show the results for the choice $\lambda =  \lambda_{\mathrm{MSA}}$ in the case of NP+LEMC for the large concentration of $c=1$ M where $\lambda_{\mathrm{D}}$ and $\lambda_{\mathrm{MSA}}$ are quite different ($0.304$ and $0.415$, respectively).
As it was already pointed out in Ref.~\cite{fertig_jpcc_2019}, using $\lambda_{\mathrm{MSA}}$ for NP+LEMC and $\lambda =\lambda_{\mathrm{D}}$ for PNP produces better agreement between the two models. 
This is also seen here: the NP+LEMC and PNP curves agree if we use the appropriate screening lengths for the two methods.

Fundamentally, $\lambda_{\mathrm{MSA}}$ suits LEMC better because MSA also takes BMF correlations into account even if in an approximate manner.
We use $\lambda_{\mathrm{D}}$ in this work because it is easier to calculate, and using it also makes our point.
The equations for $\lambda_{\mathrm{MSA}}$ are found in earlier publications.~\cite{blum_mp_1975,blum_jcp_1977,nonner_bj_2000,fertig_jpcc_2019}

The PNP curve in Fig.~\ref{fig3:scaling}A has also been computed by fixing $\sigma$ and scanning $U$.
The resulting curve practically coincides with the one shown in the figure for fixed $U$ and scanned $\sigma$.
This implies that scaling works better in the mean-field approximation, while ionic correlations may cause deviations in the NP+LEMC data.
This is more apparent for multivalent electrolytes where ionic correlations are strong (data not shown).~\cite{fertig_pccp_2020}

If we want to answer the question why changing $\sigma$, $H$, or $U$ has similar effects on rectification for a given value of $c$ and $R$, we need to examine more detailed results provided by the simulations, e.g., concentration, electrical potential, and electrochemical potential profiles.

\subsection{Self consistent NP+LEMC system}

Cation and anion concentration profiles are separated both in the radial and axial dimensions.
In the radial dimension, it is the surface charge that separates the cation and anion profiles and produces the radial DL.
The overlap of this DL in the centerline of the pore produces the depletion zone that is tuned by the applied field.
Here, we do not show radial profiles; they have been discussed in detail in our previous publications, for example, in Figs.\ 2 and 5 of Ref.~\cite{fertig_jpcc_2019} and Figs.\ 5, 6, and 9 of Ref.~\cite{fertig_pccp_2020}.

In the axial dimension, the axially asymmetric pore charge and the applied field modifies the cationic and anionic distributions in such a way that it produces a charge imbalance along the $z$ axis. 
We call this charge imbalance in the $z$-dimension an axial DL.
If the screening length is large, the radial and axial DLs overlap and their structures are mutually correlated.

An accurate screening, however, seems to be based on a decoupling of the radial and axial effects.
If the pore is short compared to $\lambda_{\mathrm{D}}$ ($\lambda_{\mathrm{D}} /H$ is large), the ``p'' region has a strong effect on the ionic distributions in the ``n'' region, and vice versa.
The neighboring region influences the radial behavior of the concentration profiles, the degree of DL overlap, and, thus, the formation of depletion zones.

\begin{figure*}
\centering
\includegraphics[width=0.6\textwidth]{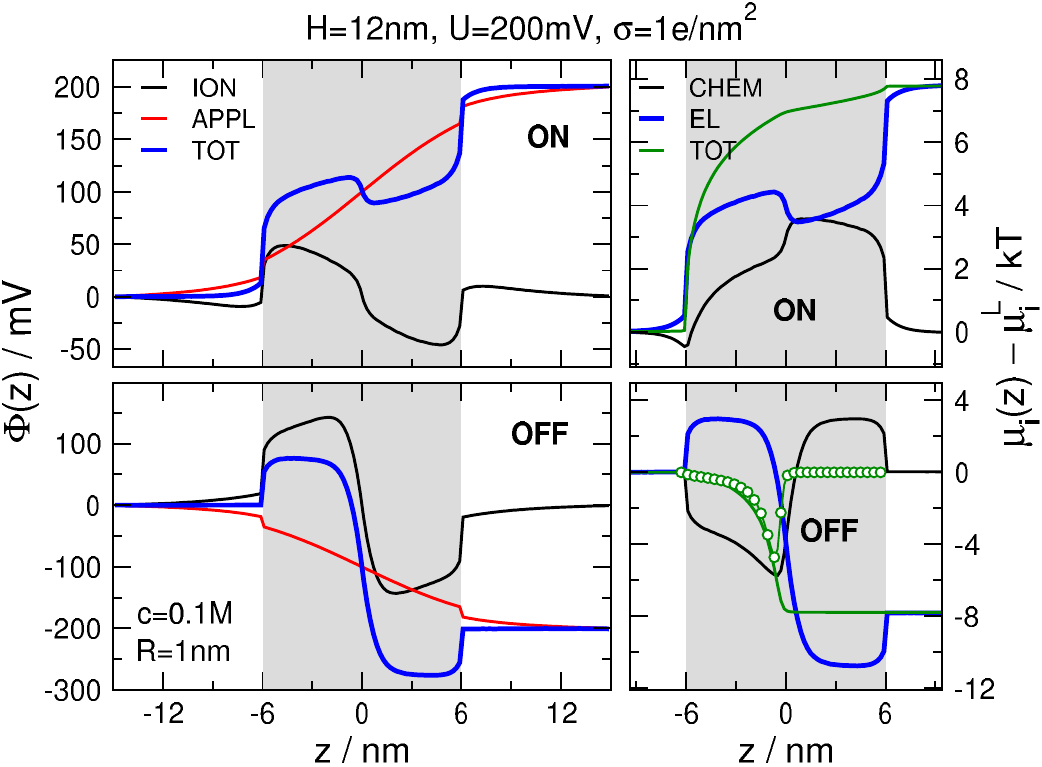}
\caption{Cross-section-averaged axial profiles for the mean electrical potential ($\Phi(z)$, left) and the electrochemical potential of the cation ($\mu_{+}(z)$, right).
Top and bottom rows refer to the ON and OFF voltages ($U=200$ and $-200$ mV), respectively.
The mean electrical potential (blue) is split into the terms of the applied potential (red) and the potential produced by the ions and the pore charges (black) as defined in Eq.~\ref{eq:potential-terms}.
The electrochemical potential (green) is split into the chemical (black) and electrical (blue) terms as defined in Eqs.~\ref{eq:elchempot}-\ref{eq:elterm}.
The blue curves on the right are equal to the blue curves on the left multiplied by $e/kT$.
The $\mu_{i}^{\mathrm{CHEM}}(z)$ and $\mu_{i}(z)$ terms are shifted with $\mu_{i}^{\mathrm{L}}$.
The open green symbols represent the derivative of the electrochemical potential, $\mathrm{d}\mu_{+}(z)/\mathrm{d}z$.
The parameters are $R=1$ nm, $H=12$ nm, $c=0.1$ M, and $\sigma=1$ $e$/nm$^{2}$.
}
\label{fig4:chempots}
\end{figure*}

Fig.~\ref{fig4:chempots} explains how $c_{i}(z)$, $\Phi(z)$, and $\mu_{i}(z)$ cooperate to provide a self consistent solution of the NP+LEMC system (and the PNP system as well).
The top and bottom rows show the ON and OFF states ($U=200$ and $-200$ mV), respectively.
The left panels show the cross-section-averaged (Eq.~\ref{eq:cz}) mean electrical potential profiles (blue) and their two components:
\begin{equation}
 \Phi(z)= \Phi^{\mathrm{APPL}}(z) + \Phi^{\mathrm{ION}}(z),
 \label{eq:potential-terms}
\end{equation} 
where $\Phi^{\mathrm{APPL}}(z)$ is the applied electrical potential (red) obtained by solving Laplace's equation (Eq.~\ref{eq:laplace}) with the prescribed Dirichlet boundary conditions ($\Phi=0$ on the left end, while $\Phi = U$ on the right end), while $\Phi^{\mathrm{ION}}(z)$ is the mean electrical potential produced by the ions and the surface charges inside the pore (black).
$\Phi^{\mathrm{ION}}(z)$ has been computed by ``on the fly'' sampling in the LEMC simulations, but solving Poisson's equation leads to the same result.
A potential profile with positive and negative peaks in the ``p'' and ``n'' regions is produced by the procedure.

The total mean electrical potential, $\Phi(z)$, multiplied by $z_{i}e$ is defined as the electrical part, $\mu_{i}^{\mathrm{EL}}(z)$, of the electrochemical potential, $\mu_{i}(z)$ (see Eqs.~\ref{eq:elchempot} and \ref{eq:elterm}), while the difference between them is defined as the chemical term, $\mu_{i}^{\mathrm{CHEM}}(z)$ (see Eq.~\ref{eq:chempot}).
The $\mu_{i}^{\mathrm{CHEM}}(z)$ and $\mu_{i}^{\mathrm{EL}}(z)$ contributions are plotted by black and blue lines, while $\mu_{i}(z)$ is plotted by green lines in the right panels of Fig.~\ref{fig4:chempots}.
Because the gradient of $\mu_{i}$ is the driving force of the transport, we show the derivative of $\mu_{i}(z)$ with open symbols in the bottom-right panel to illustrate that $\mathrm{d}\mu_{i}(z)/\mathrm{d}z$ has a minimum where $\mu_{i}^{\mathrm{CHEM}}(z)$ profile has a minimum.

The minimum of the $\mu_{i}^{\mathrm{CHEM}}(z)$ profile, on the other hand, corresponds to a minimum in the $c_{i}(z)$ profile (a depletion zone) because $\mu_{i}^{\mathrm{CHEM}}(z) \sim \ln c_{i}(z)$. 
This is because the BMF term in Eq.~\ref{eq:chempot} is small for the 1:1 system studied here. 

In general, $\mu_{i}(z)$ (green curve) must be monotonic, so $\mu_{i}^{\mathrm{CHEM}}(z)$ must balance $\mu_{i}^{\mathrm{EL}}(z)$ accordingly. 
The result is that depletion zones are formed in the OFF state at the junction (the interface of the ``p'' and ``n'' region).

In summary, in the self consistent calculation of the NP+LEMC system, ionic distribution profiles, $c_{i}(z,r)$, must be obtained that produce a mean electrical potential, $\Phi(z,r)$, that, together with $\mu_{i}^{\mathrm{CHEM}}(z,r)$, produce a $\mu_{i}(z,r)$ profile that, together with $c_{i}(z,r)$, produce a flux density, $\mathbf{j}_{i}(z,r)$, that satisfies conservation of mass, $\nabla\cdot \mathbf{j}_{i}(z,r)=0$.  

\subsection{Slope conductance approach}

Plotting concentration profiles is not necessarily the best way to get the appropriate insight in the case of nanopores, especially, if the nanopore's behavior is driven by depletion zones as it was illustrated by Fig.~\ref{fig2:c}.

It has already been realized and discussed in several publications of ours \cite{gillespie_bj_2008_ca,gillespie_bj_2008_nanopore,he_jacs_2009,boda_jgp_2009,malasics_bba_2010_trivalent,valisko_jcp_2019,fertig_pccp_2020,boda_entropy_2020} that the conductance of the pore is better described by the reciprocal of the concentration than the concentration itself as soon as the resistance of the pore is determined by depletion zones along the $z$-axis.  
Depletion zones are high-resistance elements in the consecutive segments of the pore along the $z$-axis that are considered as resistors connected in series. 
This approach works if there are so deep depletion zones in the pore that the importance of the peaks are dwarfed compared to them.

This is the case, for example, in ion channels that are naturally narrow pores where coions are excluded from regions dominated by charged amino acids as in the case of calcium channels \cite{gillespie_bj_2008_ca,boda_jgp_2009,malasics_bba_2010_trivalent} whose selectivity filter is lined by E or D amino acids.
Coions, however, can also be excluded from wide pores if the screening length of the electrolyte is large enough compared to the radius of the pore.~\cite{madai_pccp_2018,fertig_jpcc_2019,valisko_jcp_2019,fertig_pccp_2020,boda_entropy_2020,sarkadi_jcp_2021,levine_jcis_1975,balme_sr_2015,uematsu_jpcb_2018,green_jcp_2021}

An elegant quantification of this idea is the slope conductance approach \cite{gillespie_bj_2008_ca,gillespie_bj_2008_nanopore,boda_jgp_2009} in which we assume that the chemical potential is constant in the radial dimension ($\mu_{i}(z,r)\approx \mu_{i}(z)$) inside the pore.
The current carried by an ionic species $i$, $I_{i}$, is computed from the cross-sectional integral of $\mathbf{j}_{i}(z,r)$, inside the pore.
By substituting $j_{i}(z,r)$ from the NP equation (Eq.\ \ref{eq:np}) we obtain that
\begin{eqnarray}
 I_{i} & \approx & -\dfrac{z_{i}eD_{i}^{\mathrm{pore}}}{kT}  \dfrac{\mathrm{d}\mu_{i}(z)}{\mathrm{d}z} \int\limits_{0}^{R-d_{i}/2} c_{i}(z,r) 2\pi r \mathrm{d} r  = \nonumber \\
 & = & -\dfrac{z_{i}eD_{i}^{\mathrm{pore}}}{kT} \dfrac{\mathrm{d}\mu_{i}(z)}{\mathrm{d}z} A_{i} c_{i}(z) ,
 \label{eq:I-cz}
\end{eqnarray} 
where $A_{i}=(R-d_{i}/2)^{2}\pi$ is the effective cross section (where $c_{i}(z,r)$ is nonzero for $-H/2<z<H/2$) and $c_{i}(z)$ is the cross-section-averaged axial concentration profile defined in Eq.~\ref{eq:cz}. 
If we divide by $c_{i}(z)$ and integrate, Eq.\ \ref{eq:I-cz} can be expressed as
\begin{equation}
 g_{i} = \dfrac{I_{i}}{U} = -\dfrac{z_{i}^{2}e^{2}A_{i}D_{i}^{\mathrm{pore}} }{kT} \left( \int\limits_{-H/2}^{H/2} \dfrac{\mathrm{d}z}{c_{i}(z)} \right)^{-1},
 \label{eq:resistance}
\end{equation}
where $g_{i}$ is the conductance for ionic species $i$ and 
\begin{eqnarray}
 U &=& \frac{1}{z_{i}e} \int_{-H/2}^{H/2} \mathrm{d} \mu_{i}(z) \nonumber \\
 &=& \frac{1}{z_{i}e} \left[ \mu_{i}(H/2) {-} \mu_{i}(-H/2) \right] \nonumber \\
 &\approx& \Phi(H/2) {-} \Phi(-H/2)
 \label{eq:u-intmu}
\end{eqnarray} 
is the potential drop across the pore. 
Eq.~\ref{eq:u-intmu} assumed that the concentration is the same on the two sides of the membrane, so only the $\mu_{i}^{\mathrm{EL}}(z)=z_{i}e\Phi(z)$ term of $\mu_{i}$ is different at $z=-H/2$ and $H/2$.

The conductance, therefore, is associated with the reciprocal of the integral of $c_{i}^{-1}(z)$ along the pore.
If $c_{i}(z)$ is very small, $c_{i}^{-1}(z)$ is very large, its integral is large, and the conductance is small (resistance large).

\begin{figure}
\centering
\includegraphics[width=0.35\textwidth]{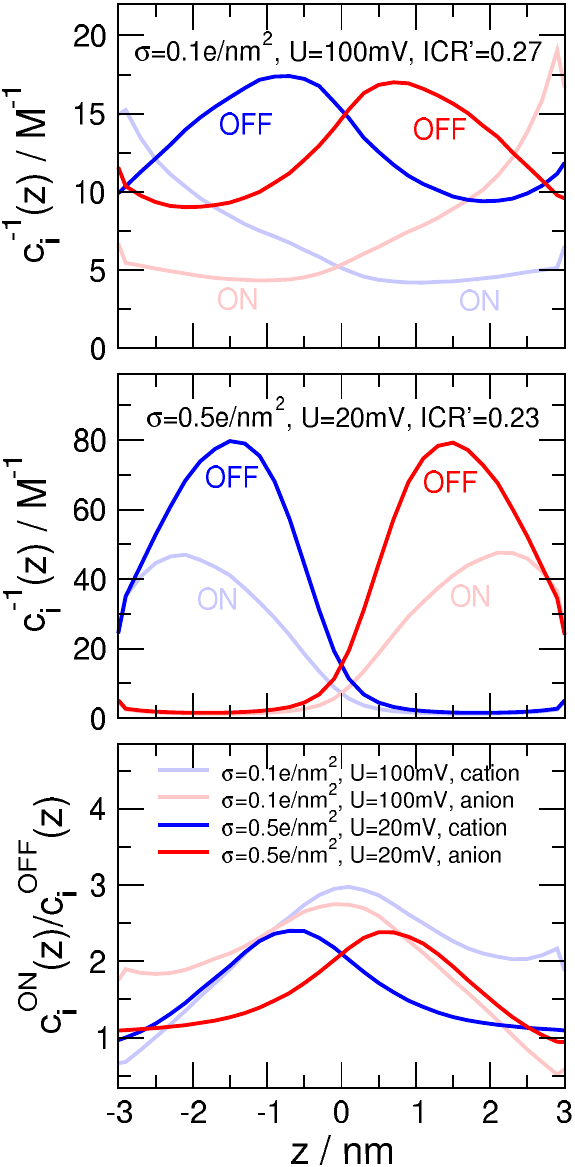}
\caption{Axial reciprocal concentration profiles are shown for the same state point shown in Fig.~\ref{fig2:c} in the top and the middle panels. 
The surface charge density and the voltage are $\sigma=0.1$ $e$/nm$^{2}$ and $U=100$ mV in the top panel, while they are $\sigma=0.5$ $e$/nm$^{2}$ and $U=20$ mV in the bottom panel (their product is fixed).
Blue and red lines refer to cation and anion profiles, respectively.
Red and light red (blue and light blue) profiles refer to OFF and ON states, respectively.
The bottom panel shows the ratio of the ON and OFF concentration profiles.
} 
\label{fig5:invc}
\end{figure}

Because the radii and diffusion constants of the cations and anions are the same in this study, their currents are similar, and we can write for the rectification that
\begin{eqnarray}
 \mathrm{ICR} \sim \frac{I^{\mathrm{ON}}_{i}}{I^{\mathrm{OFF}}_{i}} & \sim & \dfrac{\left[ \int \left(c_{i}^{\mathrm{ON}}(z)\right)^{-1} dz \right]^{-1}}{\left[ \int \left(c_{i}^{\mathrm{OFF}}(z)\right)^{-1} dz \right]^{-1}} \nonumber \\
 &=&  \dfrac{ \int \left(c_{i}^{\mathrm{OFF}}(z)\right)^{-1} dz }{ \int \left(c_{i}^{\mathrm{ON}}(z)\right)^{-1} dz }, 
\end{eqnarray} 
that is, rectification is proportional to the ratio of the integrals of the reciprocal concentration profiles in the OFF and ON states for either $i$.
This is shown in Fig.~\ref{fig5:invc} for the cases already depicted in Fig.~\ref{fig2:c}.

\begin{figure*}
\centering
\includegraphics[width=0.6\textwidth]{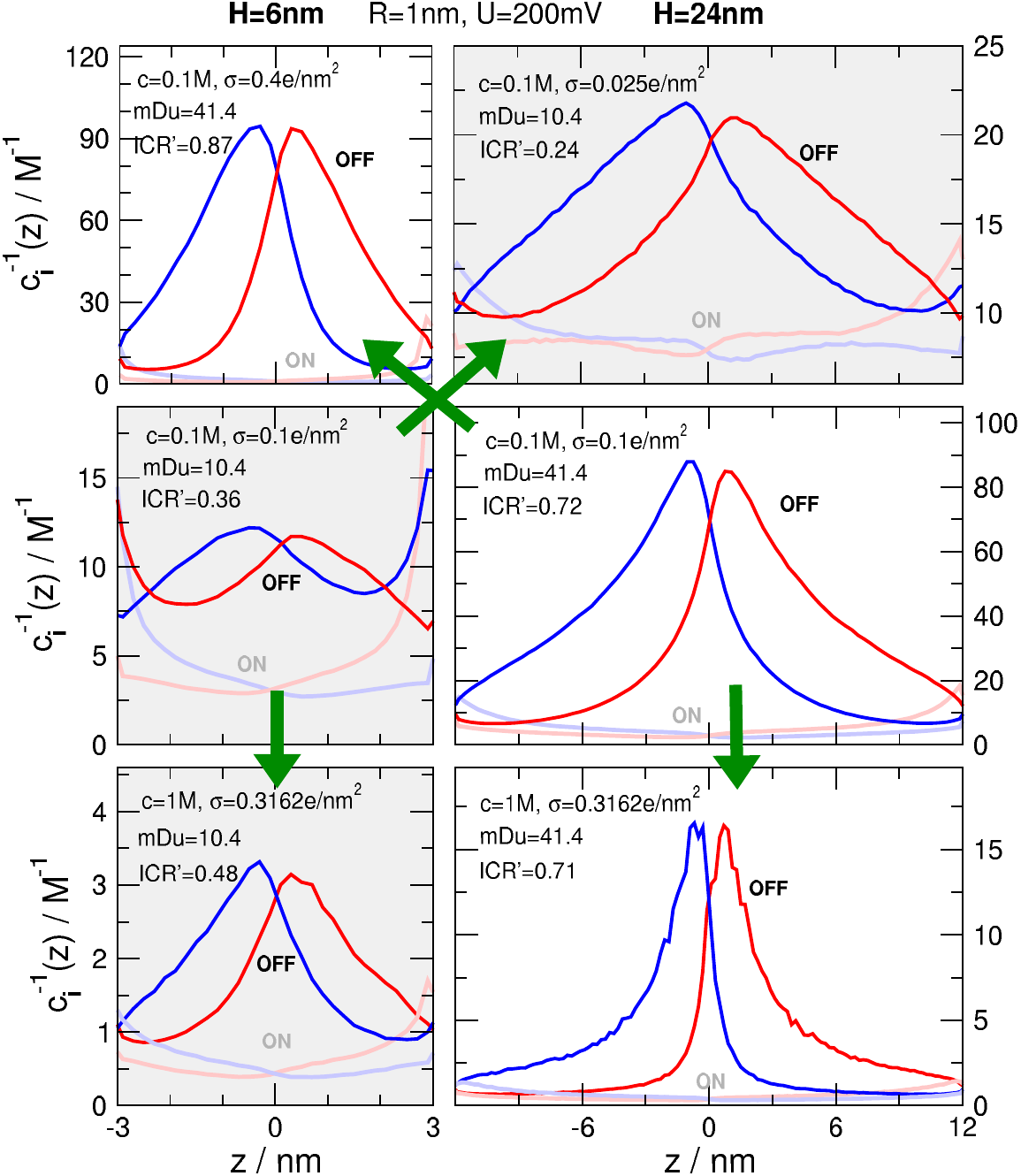}
\caption{Axial reciprocal concentration profiles for fixed pore radius, $R=1$ nm, and voltage, $U=200$ mV.  The pore length is $H=6$ nm on the left hand side, while it is $H=24$ nm on the right hand side. 
The surface charge density, $\sigma$, and concentration, $c$, are chosen such a way in the different panels that the mechanism of scaling is illustrated through the areas under the curves (see Eq.~\ref{eq:resistance}).
In the middle panels, the base point is $\sigma=0.1$ $e$/nm$^{2}$ and $c=0.1$ M. 
In the top-left panel, $H$ is decreased to $6$ nm, while $\sigma$ is increased to $0.4$ $e$/nm$^{2}$ so that mDu is the same as in the middle-right panel.
In the top-right panel, $H$ is increased to $24$ nm, while $\sigma$ is decreased to $0.025$ $e$/nm$^{2}$ so that mDu is the same as in the middle-left panel.
In the bottom panels, $c$ is increased to $1$ M, while $\sigma$ is increased to $0.3162$ $e$/nm$^{2}$ so that mDu is the same as in the middle panels ($\sigma \lambda_{\mathrm{D}}$ is the same).
Blue and red lines refer to cation and anion profiles, respectively.
Red and light red (blue and light blue) profiles refer to OFF and ON states, respectively.
} 
\label{fig6:H-profiles}
\end{figure*}

\subsection{Concentration profiles and reciprocal concentration profiles}

\paragraph*{The effect of voltage and surface charge density}

In Fig.~\ref{fig5:invc}, peaks in the OFF profile correspond to depletion zones, so we can visualize depletion zones better this way.
The above analysis and this figure explain why Fig.~\ref{fig2:c} alone is misleading if we want to understand the mechanisms.
There, the relation between the ON and OFF curves' behaviors is so different in the top and bottom panels that it is hard to deduct from that figure why the pore rectifies similarly in the two cases.

The explanation is that Fig.~\ref{fig2:c} overemphasizes the peaks, while it is the depletion zones that are relevant for rectification. 
Looking at Fig.~\ref{fig5:invc}, the contradiction can be resolved easily.
The ratio of the integrals of the OFF and the ON profiles can be similar even if the actual behaviors of these profiles are quite different.
Indeed, if we plot the ratio of the ON and OFF concentration profiles (bottom panel), it is apparent that they are similar in the two cases.
Note that one must be careful with the $c_{i}^{\mathrm{ON}}(z)/c_{i}^{\mathrm{OFF}}(z)$ profiles because generally the integrals of these profiles do not have anything to do with the rectification.
Despite this restriction, the $c_{i}^{\mathrm{ON}}(z)/c_{i}^{\mathrm{OFF}}(z)$ function is useful in drawing conclusions for the general behavior of the device function.
If the $c_{i}^{\mathrm{ON}}(z)/c_{i}^{\mathrm{OFF}}(z)$ profile is similar in two cases, rectification may also be similar.

This is what we can see in Fig.~\ref{fig5:invc} by comparing the top and the middle panels.
The top panel refers to $\sigma=0.1$ $e$/nm$^{2}$ and $U=100$ mV; a relatively small surface charge density produces smaller peaks and less deep depletion zones, namely, counter- and coions are separated less.
A larger voltage is needed to attain a separation of the ON and OFF profiles so that we obtain a given rectification $\mathrm{ICR}'=0.27$.
The middle panel refers to the same $\sigma U$ product, but for a larger surface charge density, $\sigma=0.5$ $e$/nm$^{2}$, and a smaller voltage, $U=20$ mV.
In this case, we obtain a similar rectification $\mathrm{ICR}'=0.23$ by a larger $\sigma$ separating cation and anion profiles more, but, in turn, a smaller voltage separating the ON and OFF profiles.
The bottom panel shows the $c_{i}^{\mathrm{ON}}(z)/c_{i}^{\mathrm{OFF}}(z)$ profiles.

Although $\sigma$ and $U$ modify the ionic distributions via different mechanisms, the resulting effects are similar.
The surface charge separates cation and anion profiles; in each region there is always a peak of counterions and a depletion zone of coions.
The voltage then modulates these profiles, but it modulates them in different ways at its ON and OFF signs.
The OFF voltage makes the depletion zones deeper, while the ON voltage makes the peaks higher.
Increasing either $\sigma$ or $U$ increases the difference between the ON and OFF profiles, but via different mechanisms.

\begin{figure*}
\centering
\includegraphics[width=0.8\textwidth]{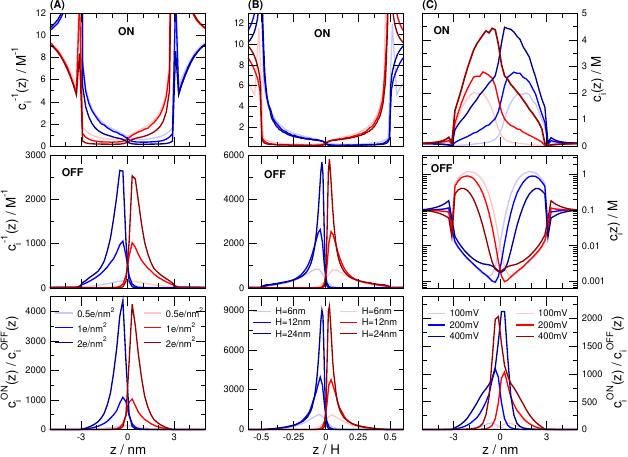}
\caption{Axial reciprocal concentration profiles (middle and top rows) and the $c_{i}^{\mathrm{ON}}(z)/c_{i}^{\mathrm{OFF}}(z)$ profiles (bottom row).
In the respective columns only one parameter is changed: (A) surface charge density, (B) pore length, (C) voltage.
The rest of the parameters are kept fixed and are $R=1$ nm, $c=0.1$ M, $H=6$ nm, $U=200$ mV, and $\sigma=1$ $e$/nm$^{2}$.
Blue and red lines refer to cation and anion profiles, respectively.
The (C) column shows the $c_{i}(z)$ profiles, while columns (A) and (B) shows the $c_{i}^{-1}(z)$ profiles.
} 
\label{fig7:sig-H-U-dep}
\end{figure*}

\paragraph*{The effect of pore length and surface charge density}
Now let us turn our attention to the interplay of $\sigma$ and $H$.
The product of these two parameters determines the net pore charge in a half region, $\pm\sigma H \pi R$, if $R$ is constant.
It is plausible that more net charge repels more coions from the pore thus producing a deeper depletion zone.

Fig.~\ref{fig6:H-profiles} shows reciprocal concentration profiles to illuminate this.
The middle row shows our two base points, where $\sigma=0.1$ $e$/nm$^{2}$ and $c=0.1$ M (voltage is kept fixed in these calculations at $U=200$ mV).
The left and right panels show the results for $H=6$ and $24$ nm, respectively.
The $H=24$ nm results refer to $4$ times larger $\mathrm{mDu}$, and, thus, a larger rectification according to scaling in Fig.~\ref{fig3:scaling}.
Indeed, this resonates with the profiles; the ratio of the areas under the OFF profiles and the ON profiles is much larger for $H=24$ nm than for $H=6$ nm.

So, what can we do if we want to get a rectification that is similar to that for the $\sigma=0.1$ $e$/nm$^{2}$, $c=0.1$ M, $H=24$ nm case (middle right panel of Fig.~\ref{fig6:H-profiles})?
One choice is to decrease $H$ to $6$ nm, but increase $\sigma$ to $0.4$ $e$/nm$^{2}$ (top left panel of Fig.~\ref{fig6:H-profiles}).
Indeed, the relation between the OFF and ON curves is similar in this panel and the middle right panel (follow the green arrow).
The other choice is to increase $\sigma$ and decrease $\lambda_{\mathrm{D}}$ so that $\sigma \lambda_{\mathrm{D}}$ (or $\sigma / \sqrt{c}$) is the same value.
The bottom right panel shows the data for $c=1$ M that corresponds to a Debye length that is $\sqrt{10}$ times smaller than in the $c=0.1$ case ($\lambda_{\mathrm{D}}=0.304$ and $0.962$ nm, respectively).
The corresponding surface charge density, therefore, is $\sqrt{10}$ times larger, namely, $\sigma=0.3162$ $e$/nm$^{2}$.
Again, the relation of the OFF and ON profiles are similar to that in the panel above (follow the green arrow).

Inversely, what can we do if we want to get a rectification that is similar to that for the $\sigma=0.1$ $e$/nm$^{2}$, $c=0.1$ M, $H=6$ nm case (middle left panel)?
One choice is to increase $H$ to $24$ nm, but decrease $\sigma$ to $0.025$ $e$/nm$^{2}$ (top right panel, follow the green arrow).
The other choice is the same as in the above paragraph: to increase $\sigma$ and decrease $\lambda_{\mathrm{D}}$ so that $\sigma \lambda_{\mathrm{D}}$ is the same value (bottom left panel, follow the green arrow).
The conclusions regarding the relations of the OFF and ON profiles are the same as in the previous paragraph.

\paragraph*{Changing voltage, pore length, and surface charge density individually}

Figs.~\ref{fig2:c}, \ref{fig5:invc}, and \ref{fig6:H-profiles} showed cases for the same mDu to reveal how and why scaling works.
Next, we show profiles to explain how and why changing a single parameter ($\sigma$, $H$, or $U$) results in a change in device behavior that is similar in the three cases.
For this, we change only one parameter while all the others are kept fixed.
This is shown in Fig.~\ref{fig7:sig-H-U-dep}.

Fig.~\ref{fig7:sig-H-U-dep}A shows the reciprocal profiles for the ON state (top panel), the OFF state (middle panel), and their ratio (bottom panel).
Blue and red colors refer to cations and anions, respectively, while different shades of blue and red refer to different surface charge densities. 
As $\sigma$ increases, the peaks in the OFF-state $c_{i}^{-1}(z)$ profiles increase.
Increasing $c_{i}^{-1}(z)$, on the other hand means deepening depletion zones.
The $c_{i}^{\mathrm{ON}}(z)/c_{i}^{\mathrm{OFF}}(z)$ profile is very similar to the OFF-state $c_{i}^{-1}(z)$ profile.
This means that rectification is chiefly determined by the behavior in the OFF state.

Fig.~\ref{fig7:sig-H-U-dep}B is similar except that now the different shades of blue and red refer to different pore lengths, $H$, and that we plot the profiles as functions of $z/H$.
Plotting this way reveals that increasing $H$ has little effect on the ON profiles, while it produces larger peaks in the OFF-state $c_{i}^{-1}(z)$ profiles, e.g., deeper depletion zones.

\begin{figure}{b!}
\centering
\includegraphics[width=0.35\textwidth]{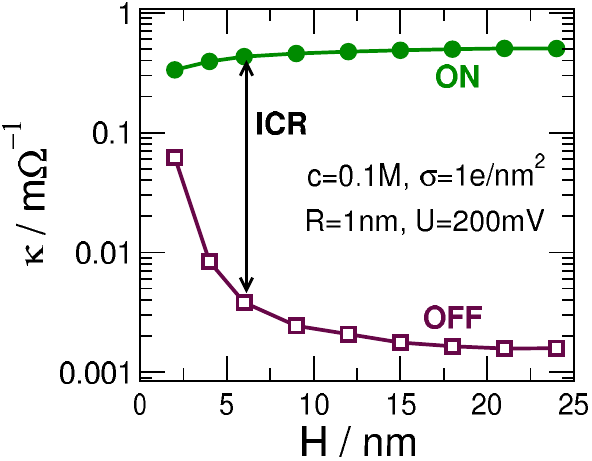}
\caption{The specific conductance of the nanopore as a function of $H$ for the parameters $R=1$ nm, $c=0.1$ M, $\sigma=1$ $e$/nm$^{2}$, and $U=200$ mV.
The two curves represent the ON and OFF states, while their distance in the figure is related to $\mathrm{ICR}=I^{\mathrm{ON}}/I^{\mathrm{OFF}}=\kappa^{\mathrm{ON}}/\kappa^{\mathrm{OFF}}$ due to the logarithmic scale of the ordinate.
Error bars are within the size of the symbols.
} 
\label{fig8:currents-Hdep}
\end{figure}

The result that rectification is rather determined by the OFF state when we change $H$ is also supported by Fig.~\ref{fig8:currents-Hdep} that shows the nanopores' specific conductances defined as 
\begin{equation}
 \kappa = \frac{IH}{UR^{2}\pi}
 \label{eq:kappa}
\end{equation} 
in the ON and OFF states.
Because $\kappa$ is plotted on a logarithmic scale, the distance between the two curves corresponds to rectification.
Rectification increases because the rate of the decrease of the OFF conductance is larger than the rate of the increase of the ON conductance.

As opposed to Figs.~\ref{fig7:sig-H-U-dep}A and B that showed reciprocal concentration profiles, Fig.~\ref{fig7:sig-H-U-dep}C shows the concentration profiles for different voltages.
This is because changing voltage rather changes the ON-state concentration profiles. 
Comparing the $c_{i}^{\mathrm{ON}}(z)/c_{i}^{\mathrm{OFF}}(z)$ profiles to either the ON or the OFF profiles, we observe that the trend showed by them (increasing peaks with increasing voltage) agrees with the trend shown by the ON-state concentration profile.

\begin{figure}
\centering
\includegraphics[width=0.35\textwidth]{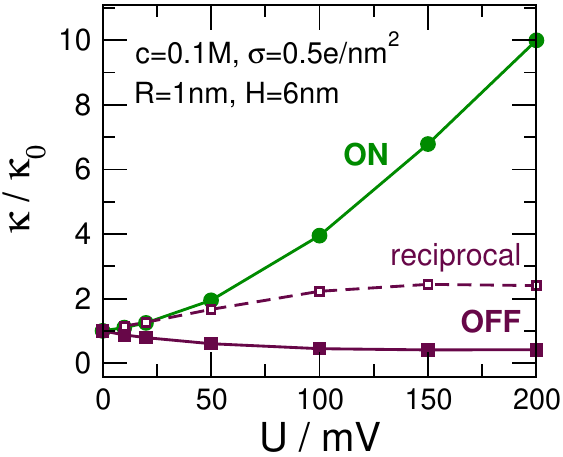}
\caption{The specific conductance of the nanopore (normalized by the $U\rightarrow 0$ value, $\kappa_{0}$) as a function of $U$ for the parameters $R=1$ nm, $H=6$ nm, $c=0.1$ M, and $\sigma=0.5$ $e$/nm$^{2}$.
The two solid curves represent the ON and OFF states, while their ratio corresponds to rectification, ICR.
For the OFF state, the dashed curve is the reciprocal of the $\kappa/\kappa_{0}$ curve.
Rectification, ICR, is the product of the ON-state curve (green) and the OFF-state reciprocal curve (maroon dashed).
Error bars are within the size of the symbols.
} 
\label{fig9:g-Udep}
\end{figure}

The result that rectification is rather determined by the ON state when we change $U$ is also supported by Fig.~\ref{fig9:g-Udep} that shows the nanopores' specific conductances in the ON and OFF states as defined in Eq.~\ref{eq:kappa}.
It is apparent that rectification increases because the rate of the increase of the ON-state specific conductance is larger than the rate of the decrease of the OFF-state specific conductance (or the OFF-state reciprocal specific conductance).

\section{Summary}
 
We showed that by a little modification of the mDu parameter (Eq.~\ref{eq:mDu_new}) introduced in our previous study \cite{sarkadi_jcp_2021}, a useful scaling parameter can be obtained for bipolar nanopores by including the voltage.
The case of bipolar nanopores (bipolar surface charge pattern) is more complicated from several points of view than the case of selective nanopores (uniform surface charge density).
The device function is rectification obtained from two separate simulation for $U$ and $-U$.
The nanotube limit, $H\rightarrow\infty$, for which we obtained an exact scaling in Ref.~\cite{sarkadi_jcp_2021}, makes sense only as a long-pore limit here.
The two oppositely charged half pores ``communicate'' with each other; the pore charge and the ionic distribution of one influences the ionic distribution of the other.

Despite this complex picture and the fact that mDu contains all the relevant parameters ($R$, $H$, $c$, $\sigma$, and $U$), we found a good, although not perfectly accurate, scaling behavior.
This behavior is useful not only from a fundamental point of view, but also from a practical point of view.
If we know, from measurements, the properties of the nanopore for a given set of parameters, we can predict it for another set of parameters as soon as scaling is valid.

Scaling, however, is not necessarily a universal feature.
Strong ionic correlations present for multivalent ions may cause deviations from the smooth monotonic scaling observed for 1:1 electrolytes. 
Therefore, the scaling behavior as reported here and in our earlier publications \cite{madai_pccp_2018,fertig_jpcc_2019,sarkadi_jcp_2021} can serve as a gold standard the deviations from which indicate the strength of ionic correlations.

\appendix
\section{Local Equilibrium Monte Carlo}
\label{sec:appendix}

Let us divide the simulation cell into subvolumes $\mathcal{B}^{\alpha}$, characterize this subvolume with the electrochemical potential, $\mu_{i}^{\alpha}$, and assume that this value is constant in $\mathcal{B}^{\alpha}$. 
We designate the $\mu_{i}^{\alpha}$ and $c_{i}^{\alpha}$ value to the mass center of the $\mathcal{B}^{\alpha}$ volume element.
We assume that the subvolumes are open systems with a constant volume ($V^{\alpha}$), temperature ($T$), and electrochemical potential ($\mu_{i}^{\alpha}$) and that they are in local equilibrium. 
In the spirit of GCMC simulations, we apply particle insertion/deletions with the acceptance probability $\min(1;p^{\alpha}_{i,\chi}(\mathbf{r}))$, where%
 \begin{equation}
  p^{\alpha}_{i,\chi}(\mathbf{r})= \dfrac{N^{\alpha}_{i}!(V^{\alpha})^{\chi}}{(N^{\alpha}_i+\chi)!} \exp\left(- \dfrac{\Delta U (\mathbf{r}) - \chi\mu^{\alpha}_i}{kT}\right).
\label{eq:valseg}
 \end{equation}
Here, $N^{\alpha}_i$ is the number of ions of type $i$ in subvolume $\mathcal{B}^{\alpha}$ before insertion/deletion, $\Delta U (\mathbf{r})$ is the change of the system's potential energy during particle insertion to position $\mathbf{r}$ (or deletion from there), $\chi=1$ for insertion, and $\chi=-1$ for deletion.
The difference between this method and equilibrium GCMC is that $\mu_{i}$ is space-dependent and the acceptance criterion is referred to a given subvolume instead of the whole simulation cell.
The effect of the surrounding of subvolume $\mathcal{B}^{\alpha}$, however, is taken into account in the simulation through the energy change that includes all the interactions from other subvolumes, not only interactions between ions in $\mathcal{B}^{\alpha}$.

Although it is tempting to consider the subvolume $\mathcal{B}^{\alpha}$ as a distinct thermodynamic system with its own ensemble of states and to  include the effect of other subvolumes as an external constraint, this is not the case.
The ensemble of states belongs to the whole system because ion configurations in subvolume $\mathcal{B}^{\alpha}$ should be collected for every possible ion configurations of all the other subvolumes. 
Therefore, the independent variables of this ensemble are $T$ and $\{ V^{\alpha}, \mu_{i}^{\alpha} \}$, where $\alpha$ and $i$ run over the volume elements and particle species, respectively. 
For comparison, the variables in global equilibrium are $T$, $V$, and $\mu_{i}$, where $V$ is the total volume and $\mu_{i}$ does not depend on space. 

We solve the NP+LEMC system iteratively.
The electrochemical potential is adjusted until conservation of mass ($\nabla \cdot \mathbf{j}_{i}(\mathbf{r})=0$) is satisfied.
The procedure can be summarized as
\begin{equation}
\mu^{\alpha}_{i}[n] \,\, \xrightarrow{\mathrm{LEMC}}  \,\,  c^{\alpha}_{i}[n] \,\,  \xrightarrow{\mathrm{NP}} \,\,  \mathbf{j}^{\alpha}_{i}[n] \,\, 
\xrightarrow{\nabla \cdot \mathbf{j}=0} 
\,\, \mu^{\alpha}_{i}[n+1] .
\label{eq:circle}
\end{equation} 
The electrochemical potentials for the next iteration, $\mu_{i}^{\alpha}[n+1]$, are computed from the results of the previous iteration, $c_{i}^{\alpha}[n]$, on the basis of the divergence-theorem (also known as Gauss-Ostrogradsky's theorem).
The continuity equation is converted to a surface integral:
\begin{equation}
 0 = \int_{\mathcal{B}^{\alpha}}\nabla\cdot\mathbf{j}_{i}(\mathbf{r})\, dV = \oint_{\mathcal{S}^{\alpha}} \mathbf{j}_{i}(\mathbf{r}) \cdot \mathbf{n}(\mathbf{r})\,da,
 \label{eq:cont2}
\end{equation}
where volume $\mathcal{B}^{\alpha}$ is bounded by surface $\mathcal{S}^{\alpha}$ and $\mathbf{n}(\mathbf{r})$ denotes the normal vector pointing outward at position $\mathbf{r}$ of the surface. 

The system is solved by digitizing the problem.
Every $\mathcal{S}^{\alpha}$ surface is divided into $\mathcal{S}^{\alpha\beta}$ elements. 
Along with these elements, $\mathcal{B}^{\alpha}$ and $\mathcal{B}^{\beta}$ are adjacent cells. 
It is assumed that the concentration, the gradient of the electrochemical potential, the flux density and the diffusion coefficient are constant on a surface $\mathcal{S}^{\alpha\beta}$. 
They are denoted by hat: $\hat{c}^{\alpha\beta}_{i}$, $\nabla \hat{\mu}_{i}^{\alpha\beta}$, $\hat{\mathbf{j}}_{i}^{\alpha\beta}$, and $\hat{D}_{i}^{\alpha\beta}$.
The $\hat{c}^{\alpha\beta}_{i}$ values are obtained from the values $c_{i}^{\alpha}$ and $c_{i}^{\beta}$ via linear interpolation.
The $\nabla \hat{\mu}_{i}^{\alpha\beta}$ values are also obtained from $\mu_{i}^{\alpha}$ and $\mu_{i}^{\beta}$ assuming linearity.

Thus the integral in Eq.~\ref{eq:cont2} for a given surface $\mathcal{S}^{\alpha}$ is replaced by a sum over the surface elements that constitute $\mathcal{S}^{\alpha}$:
\begin{equation}
 0 = \sum_{\beta, \mathcal{S}^{\alpha\beta}\in\mathcal{S}^{\alpha}} \hat{\mathbf{j}}_{i}^{\alpha\beta} \cdot \mathbf{n}^{\alpha\beta}a^{\alpha\beta},
 \label{eq:int-cont}
\end{equation}
where $a^{\alpha\beta}$ is the area of surface element $\mathcal{S}^{\alpha\beta}$ and $\mathbf{n}^{\alpha\beta}$ is the outward normal vector in the center of $\mathcal{S}^{\alpha\beta}$.
The iteration procedure is described by the following steps:
\begin{enumerate}
 \item An appropriately chosen initial set of electrochemical potentials is chosen ($\mu_i^{\alpha}[1]$; in general: $\mu_i^{\alpha}[n]$, where $[n]$ denotes the $n$th  iteration). 
 \item Using these $\mu_i^{\alpha}[n]$ parameters as inputs, LEMC simulations are performed. 
The resulting concentrations are denoted by $c_i^{\alpha}[n]$.
 \item The flux computed from the $\{ \mu_{i}^{\alpha}[n],c_{i}^{\alpha}[n]\}$ pair usually does not satisfy Eq.~\ref{eq:int-cont}.
 The next set of electrochemical potential is calculated by assuming that the $\{ \mu_{i}^{\alpha}[n+1],c_{i}^{\alpha}[n]\}$ pair \textit{does} satisfy Eq.~\ref{eq:int-cont}.
 If we write the value of $ \hat{\mathbf{j}}_{i}^{\alpha\beta}$ as given by the NP equation into Eq.~\ref{eq:int-cont}, we obtain 
\begin{equation}
  0 = \sum_{\beta} \hat{D}_{i}^{\alpha\beta}\hat{c}_{i}^{\alpha\beta}[n]\nabla \hat{\mu}_{i}^{\alpha\beta}[n+1]\cdot \mathbf{n}^{\alpha\beta}\,a^{\alpha\beta},
  \label{eq:sum}
 \end{equation}
 where $\beta$ runs over all the surface elements $ \mathcal{S}^{\alpha\beta}$ that constitute $\mathcal{S}^{\alpha}$.
 Eq.~\ref{eq:sum} is a system of linear equations; the unknown variables are denoted by {$\mu_i^{\alpha,\mathrm{CAL}}[n+1]$}, where $\mathrm{CAL}$ refers to the fact that these values come from calculations by solving \ref{eq:sum}. 
 \item To achive faster and more robust convergence in the case of large driving forces, the electrochemical potential used in the $(n+1)$th iteration is mixed from the values calculated in the $(n+1)$th iteration from Eq.~\ref{eq:sum} and the values mixed in the $n$th iteration:
\begin{eqnarray}
 \mu_i^{\alpha,\mathrm{MIX}}[n+1] & = & b_{i}\mu_i^{\alpha,\mathrm{CAL}}[n+1] \nonumber \\ 
  & + & (1-b_{i})\mu_i^{\alpha,\mathrm{MIX}}[n],
\end{eqnarray}
where $b_{i}$ is a mixing parameter that determines the ratio of mixing. 
If the parameter is close to 1, faster iteration can be achieved, however, it may result in the system fluctuating between local minima. 
Smaller $b_{i}$ values prevent this fluctuation at the price of making the convergence slower.
In this work we used values $b=0.5-0.7$.
\item The input of the $(n+1)$th LEMC simulation is $\mu_i^{\alpha,\mathrm{MIX}}[n+1]$.
\end{enumerate}

The energy change $\Delta U$ contains not only the interactions between particles (and interactions of particles with the pore), but also the interaction with an external electrical potential, $\Phi^{\mathrm{APPL}}(\mathbf{r})$.
This applied potential is calculated by solving Laplace's equation 
\begin{equation}
 \nabla^{2}\Phi^{\mathrm{APPL}}(\mathbf{r})=0
\label{eq:laplace}
\end{equation} 
for the system (inside the blue line) with the prescribed Dirichlet boundary condition on the system's boundaries ($\Phi^{\mathrm{L}}$ and $\Phi^{\mathrm{R}}$ on the left and right blue line, respectively).

In the rotationally symmetric present geometry studied here, the elementary cells are $\Delta z \times \Delta r$ rectangles in the $(z,r)$ plane.
In three-dimensional space, these correspond to concentric rings with the $z$-axis in their centers.

Because of statistical uncertainties of the LEMC simulations, the iteration does not converge to an exact value of the ionic flux density, but it fluctuates around a limiting value. 
The final solution is obtained from a running average over iterations. 
Longer LEMC simulations and more iterations result in a more reliable outcome.
In this study, we used $40-100$ iterations sampling $30$million configurations in each.

\section*{Acknowledgement(s)}

We gratefully acknowledge  the financial support of the National Research, Development and Innovation Office -- NKFIH K124353. 
We acknowledge KIFÜ for awarding us access to resource based in Hungary at Szeged.

\bibliography{nanopore,book,own}

\begin{thebibliography}{79}%
\makeatletter
\providecommand \@ifxundefined [1]{%
 \@ifx{#1\undefined}
}%
\providecommand \@ifnum [1]{%
 \ifnum #1\expandafter \@firstoftwo
 \else \expandafter \@secondoftwo
 \fi
}%
\providecommand \@ifx [1]{%
 \ifx #1\expandafter \@firstoftwo
 \else \expandafter \@secondoftwo
 \fi
}%
\providecommand \natexlab [1]{#1}%
\providecommand \enquote  [1]{``#1''}%
\providecommand \bibnamefont  [1]{#1}%
\providecommand \bibfnamefont [1]{#1}%
\providecommand \citenamefont [1]{#1}%
\providecommand \href@noop [0]{\@secondoftwo}%
\providecommand \href [0]{\begingroup \@sanitize@url \@href}%
\providecommand \@href[1]{\@@startlink{#1}\@@href}%
\providecommand \@@href[1]{\endgroup#1\@@endlink}%
\providecommand \@sanitize@url [0]{\catcode `\\12\catcode `\$12\catcode
  `\&12\catcode `\#12\catcode `\^12\catcode `\_12\catcode `\%12\relax}%
\providecommand \@@startlink[1]{}%
\providecommand \@@endlink[0]{}%
\providecommand \url  [0]{\begingroup\@sanitize@url \@url }%
\providecommand \@url [1]{\endgroup\@href {#1}{\urlprefix }}%
\providecommand \urlprefix  [0]{URL }%
\providecommand \Eprint [0]{\href }%
\providecommand \doibase [0]{http://dx.doi.org/}%
\providecommand \selectlanguage [0]{\@gobble}%
\providecommand \bibinfo  [0]{\@secondoftwo}%
\providecommand \bibfield  [0]{\@secondoftwo}%
\providecommand \translation [1]{[#1]}%
\providecommand \BibitemOpen [0]{}%
\providecommand \bibitemStop [0]{}%
\providecommand \bibitemNoStop [0]{.\EOS\space}%
\providecommand \EOS [0]{\spacefactor3000\relax}%
\providecommand \BibitemShut  [1]{\csname bibitem#1\endcsname}%
\let\auto@bib@innerbib\@empty
\bibitem [{\citenamefont {Daiguji}, \citenamefont {Oka},\ and\ \citenamefont
  {Shirono}(2005)}]{daiguji_nl_2005}%
  \BibitemOpen
  \bibfield  {author} {\bibinfo {author} {\bibfnamefont {H.}~\bibnamefont
  {Daiguji}}, \bibinfo {author} {\bibfnamefont {Y.}~\bibnamefont {Oka}}, \ and\
  \bibinfo {author} {\bibfnamefont {K.}~\bibnamefont {Shirono}},\ }\href
  {\doibase 10.1021/nl051646y} {\bibfield  {journal} {\bibinfo  {journal} {Nano
  Lett.}\ }\textbf {\bibinfo {volume} {5}},\ \bibinfo {pages} {2274} (\bibinfo
  {year} {2005})}\BibitemShut {NoStop}%
\bibitem [{\citenamefont {Constantin}\ and\ \citenamefont
  {Siwy}(2007)}]{constantin_pre_2007}%
  \BibitemOpen
  \bibfield  {author} {\bibinfo {author} {\bibfnamefont {D.}~\bibnamefont
  {Constantin}}\ and\ \bibinfo {author} {\bibfnamefont {Z.~S.}\ \bibnamefont
  {Siwy}},\ }\href {\doibase 10.1103/physreve.76.041202} {\bibfield  {journal}
  {\bibinfo  {journal} {Phys. Rev. E}\ }\textbf {\bibinfo {volume} {76}},\
  \bibinfo {eid} {041202} (\bibinfo {year} {2007})}\BibitemShut {NoStop}%
\bibitem [{\citenamefont {Karnik}\ \emph {et~al.}(2007)\citenamefont {Karnik},
  \citenamefont {Duan}, \citenamefont {Castelino}, \citenamefont {Daiguji},\
  and\ \citenamefont {Majumdar}}]{karnik_nl_2007}%
  \BibitemOpen
  \bibfield  {author} {\bibinfo {author} {\bibfnamefont {R.}~\bibnamefont
  {Karnik}}, \bibinfo {author} {\bibfnamefont {C.}~\bibnamefont {Duan}},
  \bibinfo {author} {\bibfnamefont {K.}~\bibnamefont {Castelino}}, \bibinfo
  {author} {\bibfnamefont {H.}~\bibnamefont {Daiguji}}, \ and\ \bibinfo
  {author} {\bibfnamefont {A.}~\bibnamefont {Majumdar}},\ }\href {\doibase
  10.1021/nl062806o} {\bibfield  {journal} {\bibinfo  {journal} {Nano Lett.}\
  }\textbf {\bibinfo {volume} {7}},\ \bibinfo {pages} {547} (\bibinfo {year}
  {2007})}\BibitemShut {NoStop}%
\bibitem [{\citenamefont {Vlassiouk}\ and\ \citenamefont
  {Siwy}(2007)}]{vlassiouk_nl_2007}%
  \BibitemOpen
  \bibfield  {author} {\bibinfo {author} {\bibfnamefont {I.}~\bibnamefont
  {Vlassiouk}}\ and\ \bibinfo {author} {\bibfnamefont {Z.~S.}\ \bibnamefont
  {Siwy}},\ }\href {\doibase 10.1021/nl062924b} {\bibfield  {journal} {\bibinfo
   {journal} {Nano Lett.}\ }\textbf {\bibinfo {volume} {7}},\ \bibinfo {pages}
  {552} (\bibinfo {year} {2007})}\BibitemShut {NoStop}%
\bibitem [{\citenamefont {Kalman}, \citenamefont {Vlassiouk},\ and\
  \citenamefont {Siwy}(2008)}]{kalman_am_2008}%
  \BibitemOpen
  \bibfield  {author} {\bibinfo {author} {\bibfnamefont {E.~B.}\ \bibnamefont
  {Kalman}}, \bibinfo {author} {\bibfnamefont {I.}~\bibnamefont {Vlassiouk}}, \
  and\ \bibinfo {author} {\bibfnamefont {Z.~S.}\ \bibnamefont {Siwy}},\ }\href
  {\doibase 10.1002/adma.200701867} {\bibfield  {journal} {\bibinfo  {journal}
  {Adv. Mater.}\ }\textbf {\bibinfo {volume} {20}},\ \bibinfo {pages} {293}
  (\bibinfo {year} {2008})}\BibitemShut {NoStop}%
\bibitem [{\citenamefont {Vlassiouk}, \citenamefont {Smirnov},\ and\
  \citenamefont {Siwy}(2008)}]{vlassiouk_acsnanno_2008}%
  \BibitemOpen
  \bibfield  {author} {\bibinfo {author} {\bibfnamefont {I.}~\bibnamefont
  {Vlassiouk}}, \bibinfo {author} {\bibfnamefont {S.}~\bibnamefont {Smirnov}},
  \ and\ \bibinfo {author} {\bibfnamefont {Z.}~\bibnamefont {Siwy}},\ }\href
  {\doibase 10.1021/nn800306u} {\bibfield  {journal} {\bibinfo  {journal} {ACS
  Nano}\ }\textbf {\bibinfo {volume} {2}},\ \bibinfo {pages} {1589} (\bibinfo
  {year} {2008})}\BibitemShut {NoStop}%
\bibitem [{\citenamefont {Yan}\ \emph {et~al.}(2009)\citenamefont {Yan},
  \citenamefont {Liang}, \citenamefont {Fan},\ and\ \citenamefont
  {Yang}}]{yan_nl_2009}%
  \BibitemOpen
  \bibfield  {author} {\bibinfo {author} {\bibfnamefont {R.}~\bibnamefont
  {Yan}}, \bibinfo {author} {\bibfnamefont {W.}~\bibnamefont {Liang}}, \bibinfo
  {author} {\bibfnamefont {R.}~\bibnamefont {Fan}}, \ and\ \bibinfo {author}
  {\bibfnamefont {P.}~\bibnamefont {Yang}},\ }\href {\doibase
  10.1021/nl9020123} {\bibfield  {journal} {\bibinfo  {journal} {Nano Lett.}\
  }\textbf {\bibinfo {volume} {9}},\ \bibinfo {pages} {3820} (\bibinfo {year}
  {2009})}\BibitemShut {NoStop}%
\bibitem [{\citenamefont {Cheng}\ and\ \citenamefont
  {Guo}(2009)}]{cheng_acsnano_2009}%
  \BibitemOpen
  \bibfield  {author} {\bibinfo {author} {\bibfnamefont {L.-J.}\ \bibnamefont
  {Cheng}}\ and\ \bibinfo {author} {\bibfnamefont {L.~J.}\ \bibnamefont
  {Guo}},\ }\href {\doibase 10.1021/nn8007542} {\bibfield  {journal} {\bibinfo
  {journal} {ACS Nano}\ }\textbf {\bibinfo {volume} {3}},\ \bibinfo {pages}
  {575} (\bibinfo {year} {2009})}\BibitemShut {NoStop}%
\bibitem [{\citenamefont {Nguyen}, \citenamefont {Vlassiouk},\ and\
  \citenamefont {Siwy}(2010)}]{nguyen_nt_2010}%
  \BibitemOpen
  \bibfield  {author} {\bibinfo {author} {\bibfnamefont {G.}~\bibnamefont
  {Nguyen}}, \bibinfo {author} {\bibfnamefont {I.}~\bibnamefont {Vlassiouk}}, \
  and\ \bibinfo {author} {\bibfnamefont {Z.~S.}\ \bibnamefont {Siwy}},\ }\href
  {\doibase 10.1088/0957-4484/21/26/265301} {\bibfield  {journal} {\bibinfo
  {journal} {Nanotech.}\ }\textbf {\bibinfo {volume} {21}},\ \bibinfo {pages}
  {265301} (\bibinfo {year} {2010})}\BibitemShut {NoStop}%
\bibitem [{\citenamefont {Guo}, \citenamefont {Tian},\ and\ \citenamefont
  {Jiang}(2013)}]{guo_acccr_2013}%
  \BibitemOpen
  \bibfield  {author} {\bibinfo {author} {\bibfnamefont {W.}~\bibnamefont
  {Guo}}, \bibinfo {author} {\bibfnamefont {Y.}~\bibnamefont {Tian}}, \ and\
  \bibinfo {author} {\bibfnamefont {L.}~\bibnamefont {Jiang}},\ }\href
  {\doibase 10.1021/ar400024p} {\bibfield  {journal} {\bibinfo  {journal} {Acc.
  Chem. Res.}\ }\textbf {\bibinfo {volume} {46}},\ \bibinfo {pages} {2834}
  (\bibinfo {year} {2013})}\BibitemShut {NoStop}%
\bibitem [{\citenamefont {Hat{\'{o}}}\ \emph {et~al.}(2017)\citenamefont
  {Hat{\'{o}}}, \citenamefont {Valisk{\'{o}}}, \citenamefont {Krist{\'{o}}f},
  \citenamefont {Gillespie},\ and\ \citenamefont {Boda}}]{hato_pccp_2017}%
  \BibitemOpen
  \bibfield  {author} {\bibinfo {author} {\bibfnamefont {Z.}~\bibnamefont
  {Hat{\'{o}}}}, \bibinfo {author} {\bibfnamefont {M.}~\bibnamefont
  {Valisk{\'{o}}}}, \bibinfo {author} {\bibfnamefont {T.}~\bibnamefont
  {Krist{\'{o}}f}}, \bibinfo {author} {\bibfnamefont {D.}~\bibnamefont
  {Gillespie}}, \ and\ \bibinfo {author} {\bibfnamefont {D.}~\bibnamefont
  {Boda}},\ }\href {\doibase 10.1039/c7cp01819c} {\bibfield  {journal}
  {\bibinfo  {journal} {Phys. Chem. Chem. Phys.}\ }\textbf {\bibinfo {volume}
  {19}},\ \bibinfo {pages} {17816} (\bibinfo {year} {2017})}\BibitemShut
  {NoStop}%
\bibitem [{\citenamefont {Matejczyk}\ \emph {et~al.}(2017)\citenamefont
  {Matejczyk}, \citenamefont {Valisk{\'{o}}}, \citenamefont {Wolfram},
  \citenamefont {Pietschmann},\ and\ \citenamefont
  {Boda}}]{matejczyk_jcp_2017}%
  \BibitemOpen
  \bibfield  {author} {\bibinfo {author} {\bibfnamefont {B.}~\bibnamefont
  {Matejczyk}}, \bibinfo {author} {\bibfnamefont {M.}~\bibnamefont
  {Valisk{\'{o}}}}, \bibinfo {author} {\bibfnamefont {M.-T.}\ \bibnamefont
  {Wolfram}}, \bibinfo {author} {\bibfnamefont {J.-F.}\ \bibnamefont
  {Pietschmann}}, \ and\ \bibinfo {author} {\bibfnamefont {D.}~\bibnamefont
  {Boda}},\ }\href {\doibase 10.1063/1.4978942} {\bibfield  {journal} {\bibinfo
   {journal} {J. Chem. Phys.}\ }\textbf {\bibinfo {volume} {146}},\ \bibinfo
  {pages} {124125} (\bibinfo {year} {2017})}\BibitemShut {NoStop}%
\bibitem [{\citenamefont {Fertig}\ \emph {et~al.}(2019)\citenamefont {Fertig},
  \citenamefont {Matejczyk}, \citenamefont {Valisk{\'{o}}}, \citenamefont
  {Gillespie},\ and\ \citenamefont {Boda}}]{fertig_jpcc_2019}%
  \BibitemOpen
  \bibfield  {author} {\bibinfo {author} {\bibfnamefont {D.}~\bibnamefont
  {Fertig}}, \bibinfo {author} {\bibfnamefont {B.}~\bibnamefont {Matejczyk}},
  \bibinfo {author} {\bibfnamefont {M.}~\bibnamefont {Valisk{\'{o}}}}, \bibinfo
  {author} {\bibfnamefont {D.}~\bibnamefont {Gillespie}}, \ and\ \bibinfo
  {author} {\bibfnamefont {D.}~\bibnamefont {Boda}},\ }\href {\doibase
  10.1021/acs.jpcc.9b07574} {\bibfield  {journal} {\bibinfo  {journal} {J.
  Phys. Chem. C}\ }\textbf {\bibinfo {volume} {123}},\ \bibinfo {pages} {28985}
  (\bibinfo {year} {2019})}\BibitemShut {NoStop}%
\bibitem [{\citenamefont {Blum}(1975)}]{blum_mp_1975}%
  \BibitemOpen
  \bibfield  {author} {\bibinfo {author} {\bibfnamefont {L.}~\bibnamefont
  {Blum}},\ }\href {\doibase 10.1080/00268977500103051} {\bibfield  {journal}
  {\bibinfo  {journal} {Mol. Phys.}\ }\textbf {\bibinfo {volume} {30}},\
  \bibinfo {pages} {1529} (\bibinfo {year} {1975})}\BibitemShut {NoStop}%
\bibitem [{\citenamefont {Blum}\ and\ \citenamefont
  {Hoeye}(1977)}]{blum_jcp_1977}%
  \BibitemOpen
  \bibfield  {author} {\bibinfo {author} {\bibfnamefont {L.}~\bibnamefont
  {Blum}}\ and\ \bibinfo {author} {\bibfnamefont {J.~S.}\ \bibnamefont
  {Hoeye}},\ }\href {\doibase 10.1021/j100528a019} {\bibfield  {journal}
  {\bibinfo  {journal} {J. Phys. Chem.}\ }\textbf {\bibinfo {volume} {81}},\
  \bibinfo {pages} {1311} (\bibinfo {year} {1977})}\BibitemShut {NoStop}%
\bibitem [{\citenamefont {Nonner}, \citenamefont {Catacuzzeno},\ and\
  \citenamefont {Eisenberg}(2000)}]{nonner_bj_2000}%
  \BibitemOpen
  \bibfield  {author} {\bibinfo {author} {\bibfnamefont {W.}~\bibnamefont
  {Nonner}}, \bibinfo {author} {\bibfnamefont {L.}~\bibnamefont {Catacuzzeno}},
  \ and\ \bibinfo {author} {\bibfnamefont {B.}~\bibnamefont {Eisenberg}},\
  }\href {\doibase 10.1016/s0006-3495(00)76446-0} {\bibfield  {journal}
  {\bibinfo  {journal} {Biophys. J.}\ }\textbf {\bibinfo {volume} {79}},\
  \bibinfo {pages} {1976} (\bibinfo {year} {2000})}\BibitemShut {NoStop}%
\bibitem [{\citenamefont {Sarkadi}\ \emph {et~al.}(2021)\citenamefont
  {Sarkadi}, \citenamefont {Fertig}, \citenamefont {Hat\'o}, \citenamefont
  {Valisk\'o},\ and\ \citenamefont {Boda}}]{sarkadi_jcp_2021}%
  \BibitemOpen
  \bibfield  {author} {\bibinfo {author} {\bibfnamefont {Z.}~\bibnamefont
  {Sarkadi}}, \bibinfo {author} {\bibfnamefont {D.}~\bibnamefont {Fertig}},
  \bibinfo {author} {\bibfnamefont {Z.}~\bibnamefont {Hat\'o}}, \bibinfo
  {author} {\bibfnamefont {M.}~\bibnamefont {Valisk\'o}}, \ and\ \bibinfo
  {author} {\bibfnamefont {D.}~\bibnamefont {Boda}},\ }\href {\doibase
  10.1063/5.0040593} {\bibfield  {journal} {\bibinfo  {journal} {J. Chem.
  Phys.}\ }\textbf {\bibinfo {volume} {154}},\ \bibinfo {pages} {154704}
  (\bibinfo {year} {2021})}\BibitemShut {NoStop}%
\bibitem [{\citenamefont {Albrecht}, \citenamefont {Gibb},\ and\ \citenamefont
  {Nuttall}(2013)}]{albrecht_chapter_2013}%
  \BibitemOpen
  \bibfield  {author} {\bibinfo {author} {\bibfnamefont {T.}~\bibnamefont
  {Albrecht}}, \bibinfo {author} {\bibfnamefont {T.}~\bibnamefont {Gibb}}, \
  and\ \bibinfo {author} {\bibfnamefont {P.}~\bibnamefont {Nuttall}},\ }in\
  \href {\doibase 10.1016/b978-1-4377-3473-7.00001-7} {\emph {\bibinfo
  {booktitle} {Engineered Nanopores for Bioanalytical Applications}}}\
  (\bibinfo  {publisher} {Elsevier {BV}},\ \bibinfo {year} {2013})\ pp.\
  \bibinfo {pages} {1--30}\BibitemShut {NoStop}%
\bibitem [{\citenamefont {Abgrall}\ and\ \citenamefont
  {Nguyen}(2008)}]{Abgrall_2008}%
  \BibitemOpen
  \bibfield  {author} {\bibinfo {author} {\bibfnamefont {P.}~\bibnamefont
  {Abgrall}}\ and\ \bibinfo {author} {\bibfnamefont {N.~T.}\ \bibnamefont
  {Nguyen}},\ }\href {\doibase 10.1021/ac702296u} {\bibfield  {journal}
  {\bibinfo  {journal} {Anal. Chem.}\ }\textbf {\bibinfo {volume} {80}},\
  \bibinfo {pages} {2326} (\bibinfo {year} {2008})}\BibitemShut {NoStop}%
\bibitem [{\citenamefont {Bocquet}\ and\ \citenamefont
  {Charlaix}(2010{\natexlab{a}})}]{bocquet_csr_2010}%
  \BibitemOpen
  \bibfield  {author} {\bibinfo {author} {\bibfnamefont {L.}~\bibnamefont
  {Bocquet}}\ and\ \bibinfo {author} {\bibfnamefont {E.}~\bibnamefont
  {Charlaix}},\ }\href {\doibase 10.1039/b909366b} {\bibfield  {journal}
  {\bibinfo  {journal} {Chem. Soc. Rev.}\ }\textbf {\bibinfo {volume} {39}},\
  \bibinfo {pages} {1073} (\bibinfo {year} {2010}{\natexlab{a}})}\BibitemShut
  {NoStop}%
\bibitem [{\citenamefont {Daiguji}(2010)}]{daiguji_csr_2010}%
  \BibitemOpen
  \bibfield  {author} {\bibinfo {author} {\bibfnamefont {H.}~\bibnamefont
  {Daiguji}},\ }\href {\doibase 10.1039/b820556f} {\bibfield  {journal}
  {\bibinfo  {journal} {Chem. Soc. Rev.}\ }\textbf {\bibinfo {volume} {39}},\
  \bibinfo {pages} {901} (\bibinfo {year} {2010})}\BibitemShut {NoStop}%
\bibitem [{\citenamefont {Eijkel}\ and\ \citenamefont {van~den
  Berg}(2010)}]{eijkel_csr_2010}%
  \BibitemOpen
  \bibfield  {author} {\bibinfo {author} {\bibfnamefont {J.~C.~T.}\
  \bibnamefont {Eijkel}}\ and\ \bibinfo {author} {\bibfnamefont
  {A.}~\bibnamefont {van~den Berg}},\ }\href {\doibase 10.1039/b913776a}
  {\bibfield  {journal} {\bibinfo  {journal} {Chem. Soc. Rev.}\ }\textbf
  {\bibinfo {volume} {39}},\ \bibinfo {pages} {957} (\bibinfo {year}
  {2010})}\BibitemShut {NoStop}%
\bibitem [{\citenamefont {Zangle}, \citenamefont {Mani},\ and\ \citenamefont
  {Santiago}(2010)}]{zangle_csr_2010}%
  \BibitemOpen
  \bibfield  {author} {\bibinfo {author} {\bibfnamefont {T.~A.}\ \bibnamefont
  {Zangle}}, \bibinfo {author} {\bibfnamefont {A.}~\bibnamefont {Mani}}, \ and\
  \bibinfo {author} {\bibfnamefont {J.~G.}\ \bibnamefont {Santiago}},\ }\href
  {\doibase 10.1039/b902074h} {\bibfield  {journal} {\bibinfo  {journal} {Chem.
  Soc. Rev.}\ }\textbf {\bibinfo {volume} {39}},\ \bibinfo {pages} {1014}
  (\bibinfo {year} {2010})}\BibitemShut {NoStop}%
\bibitem [{\citenamefont {M\'adai}\ \emph {et~al.}(2018)\citenamefont
  {M\'adai}, \citenamefont {Matejczyk}, \citenamefont {Dallos}, \citenamefont
  {Valisk\'o},\ and\ \citenamefont {Boda}}]{madai_pccp_2018}%
  \BibitemOpen
  \bibfield  {author} {\bibinfo {author} {\bibfnamefont {E.}~\bibnamefont
  {M\'adai}}, \bibinfo {author} {\bibfnamefont {B.}~\bibnamefont {Matejczyk}},
  \bibinfo {author} {\bibfnamefont {A.}~\bibnamefont {Dallos}}, \bibinfo
  {author} {\bibfnamefont {M.}~\bibnamefont {Valisk\'o}}, \ and\ \bibinfo
  {author} {\bibfnamefont {D.}~\bibnamefont {Boda}},\ }\href {\doibase
  10.1039/c8cp03918f} {\bibfield  {journal} {\bibinfo  {journal} {Phys. Chem.
  Chem. Phys.}\ }\textbf {\bibinfo {volume} {20}},\ \bibinfo {pages} {24156}
  (\bibinfo {year} {2018})}\BibitemShut {NoStop}%
\bibitem [{\citenamefont {Cengio}\ and\ \citenamefont
  {Pagonabarraga}(2019{\natexlab{a}})}]{dal_cengio_jcp_2019}%
  \BibitemOpen
  \bibfield  {author} {\bibinfo {author} {\bibfnamefont {S.~D.}\ \bibnamefont
  {Cengio}}\ and\ \bibinfo {author} {\bibfnamefont {I.}~\bibnamefont
  {Pagonabarraga}},\ }\href {\doibase 10.1063/1.5108723} {\bibfield  {journal}
  {\bibinfo  {journal} {J. Chem. Phys.}\ }\textbf {\bibinfo {volume} {151}},\
  \bibinfo {pages} {044707} (\bibinfo {year} {2019}{\natexlab{a}})}\BibitemShut
  {NoStop}%
\bibitem [{\citenamefont {Fertig}, \citenamefont {Valisk{\'{o}}},\ and\
  \citenamefont {Boda}(2020)}]{fertig_pccp_2020}%
  \BibitemOpen
  \bibfield  {author} {\bibinfo {author} {\bibfnamefont {D.}~\bibnamefont
  {Fertig}}, \bibinfo {author} {\bibfnamefont {M.}~\bibnamefont
  {Valisk{\'{o}}}}, \ and\ \bibinfo {author} {\bibfnamefont {D.}~\bibnamefont
  {Boda}},\ }\href {\doibase 10.1039/d0cp03237a} {\bibfield  {journal}
  {\bibinfo  {journal} {Phys. Chem. Chem. Phys.}\ }\textbf {\bibinfo {volume}
  {22}},\ \bibinfo {pages} {19033} (\bibinfo {year} {2020})}\BibitemShut
  {NoStop}%
\bibitem [{\citenamefont {Bazant}, \citenamefont {Thornton},\ and\
  \citenamefont {Ajdari}(2004)}]{bazant_pre_2004}%
  \BibitemOpen
  \bibfield  {author} {\bibinfo {author} {\bibfnamefont {M.~Z.}\ \bibnamefont
  {Bazant}}, \bibinfo {author} {\bibfnamefont {K.}~\bibnamefont {Thornton}}, \
  and\ \bibinfo {author} {\bibfnamefont {A.}~\bibnamefont {Ajdari}},\ }\href
  {\doibase 10.1103/physreve.70.021506} {\bibfield  {journal} {\bibinfo
  {journal} {Phys. Rev. E}\ }\textbf {\bibinfo {volume} {70}},\ \bibinfo
  {pages} {021506} (\bibinfo {year} {2004})}\BibitemShut {NoStop}%
\bibitem [{\citenamefont {Chu}\ and\ \citenamefont
  {Bazant}(2006)}]{chu_pre_2006}%
  \BibitemOpen
  \bibfield  {author} {\bibinfo {author} {\bibfnamefont {K.~T.}\ \bibnamefont
  {Chu}}\ and\ \bibinfo {author} {\bibfnamefont {M.~Z.}\ \bibnamefont
  {Bazant}},\ }\href {\doibase 10.1103/physreve.74.011501} {\bibfield
  {journal} {\bibinfo  {journal} {Phys. Rev. E}\ }\textbf {\bibinfo {volume}
  {74}},\ \bibinfo {pages} {011501} (\bibinfo {year} {2006})}\BibitemShut
  {NoStop}%
\bibitem [{\citenamefont {Bocquet}\ and\ \citenamefont
  {Charlaix}(2010{\natexlab{b}})}]{bocquet_chemsocrev_2010}%
  \BibitemOpen
  \bibfield  {author} {\bibinfo {author} {\bibfnamefont {L.}~\bibnamefont
  {Bocquet}}\ and\ \bibinfo {author} {\bibfnamefont {E.}~\bibnamefont
  {Charlaix}},\ }\href {\doibase 10.1039/b909366b} {\bibfield  {journal}
  {\bibinfo  {journal} {Chem. Soc. Rev.}\ }\textbf {\bibinfo {volume} {39}},\
  \bibinfo {pages} {1073} (\bibinfo {year} {2010}{\natexlab{b}})}\BibitemShut
  {NoStop}%
\bibitem [{\citenamefont {Bikerman}(1940)}]{bikerman_1940}%
  \BibitemOpen
  \bibfield  {author} {\bibinfo {author} {\bibfnamefont {J.~J.}\ \bibnamefont
  {Bikerman}},\ }\href {\doibase 10.1039/tf9403500154} {\bibfield  {journal}
  {\bibinfo  {journal} {Trans. Farad. Soc.}\ }\textbf {\bibinfo {volume}
  {35}},\ \bibinfo {pages} {154} (\bibinfo {year} {1940})}\BibitemShut
  {NoStop}%
\bibitem [{\citenamefont {Dukhin}(1993)}]{dukhin_advcollsci_1993}%
  \BibitemOpen
  \bibfield  {author} {\bibinfo {author} {\bibfnamefont {S.}~\bibnamefont
  {Dukhin}},\ }\href {\doibase 10.1016/0001-8686(93)80021-3} {\bibfield
  {journal} {\bibinfo  {journal} {Adv. Coll. Interf. Sci.}\ }\textbf {\bibinfo
  {volume} {44}},\ \bibinfo {pages} {1} (\bibinfo {year} {1993})}\BibitemShut
  {NoStop}%
\bibitem [{\citenamefont {Lyklema}\ \emph {et~al.}(1995)\citenamefont
  {Lyklema}, \citenamefont {de~Keizer}, \citenamefont {Bijsterbosch},
  \citenamefont {Fleer},\ and\ \citenamefont {(Eds.)}}]{lyklema_book_1995}%
  \BibitemOpen
  \bibfield  {author} {\bibinfo {author} {\bibfnamefont {J.~J.}\ \bibnamefont
  {Lyklema}}, \bibinfo {author} {\bibfnamefont {A.}~\bibnamefont {de~Keizer}},
  \bibinfo {author} {\bibfnamefont {B.}~\bibnamefont {Bijsterbosch}}, \bibinfo
  {author} {\bibfnamefont {G.}~\bibnamefont {Fleer}}, \ and\ \bibinfo {author}
  {\bibfnamefont {M.~C.~S.}\ \bibnamefont {(Eds.)}},\ }\href
  {http://gen.lib.rus.ec/book/index.php?md5=145e94c05719008b3b7f73e2a5da6e89}
  {\emph {\bibinfo {title} {Solid-Liquid Interfaces}}},\ Fundamentals of
  Interface and Colloid Science 2\ (\bibinfo  {publisher} {Elsevier, Academic
  Press},\ \bibinfo {year} {1995})\BibitemShut {NoStop}%
\bibitem [{\citenamefont {Khair}\ and\ \citenamefont
  {Squires}(2008)}]{khair_jfm_2008}%
  \BibitemOpen
  \bibfield  {author} {\bibinfo {author} {\bibfnamefont {A.~S.}\ \bibnamefont
  {Khair}}\ and\ \bibinfo {author} {\bibfnamefont {T.~M.}\ \bibnamefont
  {Squires}},\ }\href {\doibase 10.1017/s002211200800390x} {\bibfield
  {journal} {\bibinfo  {journal} {J. Fluid Mech.}\ }\textbf {\bibinfo {volume}
  {615}},\ \bibinfo {pages} {323} (\bibinfo {year} {2008})}\BibitemShut
  {NoStop}%
\bibitem [{\citenamefont {Das}\ and\ \citenamefont
  {Chakraborty}(2010)}]{das_langmuir_2010}%
  \BibitemOpen
  \bibfield  {author} {\bibinfo {author} {\bibfnamefont {S.}~\bibnamefont
  {Das}}\ and\ \bibinfo {author} {\bibfnamefont {S.}~\bibnamefont
  {Chakraborty}},\ }\href {\doibase 10.1021/la1009237} {\bibfield  {journal}
  {\bibinfo  {journal} {Langmuir}\ }\textbf {\bibinfo {volume} {26}},\ \bibinfo
  {pages} {11589} (\bibinfo {year} {2010})}\BibitemShut {NoStop}%
\bibitem [{\citenamefont {Lee}\ \emph {et~al.}(2012)\citenamefont {Lee},
  \citenamefont {Joly}, \citenamefont {Siria}, \citenamefont {Biance},
  \citenamefont {Fulcrand},\ and\ \citenamefont {Bocquet}}]{lee_nanolett_2012}%
  \BibitemOpen
  \bibfield  {author} {\bibinfo {author} {\bibfnamefont {C.}~\bibnamefont
  {Lee}}, \bibinfo {author} {\bibfnamefont {L.}~\bibnamefont {Joly}}, \bibinfo
  {author} {\bibfnamefont {A.}~\bibnamefont {Siria}}, \bibinfo {author}
  {\bibfnamefont {A.-L.}\ \bibnamefont {Biance}}, \bibinfo {author}
  {\bibfnamefont {R.}~\bibnamefont {Fulcrand}}, \ and\ \bibinfo {author}
  {\bibfnamefont {L.}~\bibnamefont {Bocquet}},\ }\href {\doibase
  10.1021/nl301412b} {\bibfield  {journal} {\bibinfo  {journal} {Nano Lett.}\
  }\textbf {\bibinfo {volume} {12}},\ \bibinfo {pages} {4037} (\bibinfo {year}
  {2012})}\BibitemShut {NoStop}%
\bibitem [{\citenamefont {Yeh}\ \emph {et~al.}(2014)\citenamefont {Yeh},
  \citenamefont {Wang}, \citenamefont {Chang},\ and\ \citenamefont
  {Yang}}]{yeh_ijc_2014}%
  \BibitemOpen
  \bibfield  {author} {\bibinfo {author} {\bibfnamefont {H.-C.}\ \bibnamefont
  {Yeh}}, \bibinfo {author} {\bibfnamefont {M.}~\bibnamefont {Wang}}, \bibinfo
  {author} {\bibfnamefont {C.-C.}\ \bibnamefont {Chang}}, \ and\ \bibinfo
  {author} {\bibfnamefont {R.-J.}\ \bibnamefont {Yang}},\ }\href {\doibase
  10.1002/ijch.201400079} {\bibfield  {journal} {\bibinfo  {journal} {Israel J.
  Chem.}\ }\textbf {\bibinfo {volume} {54}},\ \bibinfo {pages} {1533} (\bibinfo
  {year} {2014})}\BibitemShut {NoStop}%
\bibitem [{\citenamefont {Ma}\ \emph {et~al.}(2017)\citenamefont {Ma},
  \citenamefont {Guo}, \citenamefont {Jia},\ and\ \citenamefont
  {Xie}}]{ma_acssens_2017}%
  \BibitemOpen
  \bibfield  {author} {\bibinfo {author} {\bibfnamefont {Y.}~\bibnamefont
  {Ma}}, \bibinfo {author} {\bibfnamefont {J.}~\bibnamefont {Guo}}, \bibinfo
  {author} {\bibfnamefont {L.}~\bibnamefont {Jia}}, \ and\ \bibinfo {author}
  {\bibfnamefont {Y.}~\bibnamefont {Xie}},\ }\href {\doibase
  10.1021/acssensors.7b00793} {\bibfield  {journal} {\bibinfo  {journal} {{ACS}
  Sensors}\ }\textbf {\bibinfo {volume} {3}},\ \bibinfo {pages} {167} (\bibinfo
  {year} {2017})}\BibitemShut {NoStop}%
\bibitem [{\citenamefont {Xiong}\ \emph {et~al.}(2019)\citenamefont {Xiong},
  \citenamefont {Zhang}, \citenamefont {Jiang}, \citenamefont {Yu},\ and\
  \citenamefont {Mao}}]{xiong_scc_2019}%
  \BibitemOpen
  \bibfield  {author} {\bibinfo {author} {\bibfnamefont {T.}~\bibnamefont
  {Xiong}}, \bibinfo {author} {\bibfnamefont {K.}~\bibnamefont {Zhang}},
  \bibinfo {author} {\bibfnamefont {Y.}~\bibnamefont {Jiang}}, \bibinfo
  {author} {\bibfnamefont {P.}~\bibnamefont {Yu}}, \ and\ \bibinfo {author}
  {\bibfnamefont {L.}~\bibnamefont {Mao}},\ }\href {\doibase
  10.1007/s11426-019-9526-4} {\bibfield  {journal} {\bibinfo  {journal} {Sci.
  China Chem.}\ }\textbf {\bibinfo {volume} {62}},\ \bibinfo {pages} {1346}
  (\bibinfo {year} {2019})}\BibitemShut {NoStop}%
\bibitem [{\citenamefont {Poggioli}, \citenamefont {Siria},\ and\ \citenamefont
  {Bocquet}(2019)}]{poggioli_jpcb_2019}%
  \BibitemOpen
  \bibfield  {author} {\bibinfo {author} {\bibfnamefont {A.~R.}\ \bibnamefont
  {Poggioli}}, \bibinfo {author} {\bibfnamefont {A.}~\bibnamefont {Siria}}, \
  and\ \bibinfo {author} {\bibfnamefont {L.}~\bibnamefont {Bocquet}},\ }\href
  {\doibase 10.1021/acs.jpcb.8b11202} {\bibfield  {journal} {\bibinfo
  {journal} {J. Phys. Chem. B}\ }\textbf {\bibinfo {volume} {123}},\ \bibinfo
  {pages} {1171} (\bibinfo {year} {2019})}\BibitemShut {NoStop}%
\bibitem [{\citenamefont {Cengio}\ and\ \citenamefont
  {Pagonabarraga}(2019{\natexlab{b}})}]{dalcengio_jcp_2019}%
  \BibitemOpen
  \bibfield  {author} {\bibinfo {author} {\bibfnamefont {S.~D.}\ \bibnamefont
  {Cengio}}\ and\ \bibinfo {author} {\bibfnamefont {I.}~\bibnamefont
  {Pagonabarraga}},\ }\href {\doibase 10.1063/1.5108723} {\bibfield  {journal}
  {\bibinfo  {journal} {J. Chem. Phys.}\ }\textbf {\bibinfo {volume} {151}},\
  \bibinfo {pages} {044707} (\bibinfo {year} {2019}{\natexlab{b}})}\BibitemShut
  {NoStop}%
\bibitem [{\citenamefont {Kavokine}, \citenamefont {Netz},\ and\ \citenamefont
  {Bocquet}(2020)}]{kavokine_annualrev_2020}%
  \BibitemOpen
  \bibfield  {author} {\bibinfo {author} {\bibfnamefont {N.}~\bibnamefont
  {Kavokine}}, \bibinfo {author} {\bibfnamefont {R.~R.}\ \bibnamefont {Netz}},
  \ and\ \bibinfo {author} {\bibfnamefont {L.}~\bibnamefont {Bocquet}},\ }\href
  {\doibase 10.1146/annurev-fluid-071320-095958} {\bibfield  {journal}
  {\bibinfo  {journal} {Annu. Rev. Fluid Mech.}\ }\textbf {\bibinfo {volume}
  {53}} (\bibinfo {year} {2020}),\
  10.1146/annurev-fluid-071320-095958}\BibitemShut {NoStop}%
\bibitem [{\citenamefont {Noh}\ and\ \citenamefont
  {Aluru}(2020)}]{noh_acsnano_2020}%
  \BibitemOpen
  \bibfield  {author} {\bibinfo {author} {\bibfnamefont {Y.}~\bibnamefont
  {Noh}}\ and\ \bibinfo {author} {\bibfnamefont {N.~R.}\ \bibnamefont
  {Aluru}},\ }\href {\doibase 10.1021/acsnano.0c04453} {\bibfield  {journal}
  {\bibinfo  {journal} {{ACS} Nano}\ }\textbf {\bibinfo {volume} {14}},\
  \bibinfo {pages} {10518} (\bibinfo {year} {2020})}\BibitemShut {NoStop}%
\bibitem [{\citenamefont {Levy}, \citenamefont {de~Souza},\ and\ \citenamefont
  {Bazant}(2020)}]{levy_jcis_2020}%
  \BibitemOpen
  \bibfield  {author} {\bibinfo {author} {\bibfnamefont {A.}~\bibnamefont
  {Levy}}, \bibinfo {author} {\bibfnamefont {J.~P.}\ \bibnamefont {de~Souza}},
  \ and\ \bibinfo {author} {\bibfnamefont {M.~Z.}\ \bibnamefont {Bazant}},\
  }\href {\doibase 10.1016/j.jcis.2020.05.109} {\bibfield  {journal} {\bibinfo
  {journal} {J. Coll. Inter. Sci.}\ }\textbf {\bibinfo {volume} {579}},\
  \bibinfo {pages} {162} (\bibinfo {year} {2020})}\BibitemShut {NoStop}%
\bibitem [{\citenamefont {Siwy}\ \emph {et~al.}(2004)\citenamefont {Siwy},
  \citenamefont {Heins}, \citenamefont {Harrell}, \citenamefont {Kohli},\ and\
  \citenamefont {Martin}}]{Siwy_2004}%
  \BibitemOpen
  \bibfield  {author} {\bibinfo {author} {\bibfnamefont {Z.}~\bibnamefont
  {Siwy}}, \bibinfo {author} {\bibfnamefont {E.}~\bibnamefont {Heins}},
  \bibinfo {author} {\bibfnamefont {C.~C.}\ \bibnamefont {Harrell}}, \bibinfo
  {author} {\bibfnamefont {P.}~\bibnamefont {Kohli}}, \ and\ \bibinfo {author}
  {\bibfnamefont {C.~R.}\ \bibnamefont {Martin}},\ }\href {\doibase
  10.1021/ja047675c} {\bibfield  {journal} {\bibinfo  {journal} {J. Am. Chem.
  Soc.}\ }\textbf {\bibinfo {volume} {126}},\ \bibinfo {pages} {10850}
  (\bibinfo {year} {2004})}\BibitemShut {NoStop}%
\bibitem [{\citenamefont {Garaj}\ \emph {et~al.}(2010)\citenamefont {Garaj},
  \citenamefont {Hubbard}, \citenamefont {Reina}, \citenamefont {Kong},
  \citenamefont {Branton},\ and\ \citenamefont {Golovchenko}}]{garaj_n_2010}%
  \BibitemOpen
  \bibfield  {author} {\bibinfo {author} {\bibfnamefont {S.}~\bibnamefont
  {Garaj}}, \bibinfo {author} {\bibfnamefont {W.}~\bibnamefont {Hubbard}},
  \bibinfo {author} {\bibfnamefont {A.}~\bibnamefont {Reina}}, \bibinfo
  {author} {\bibfnamefont {J.}~\bibnamefont {Kong}}, \bibinfo {author}
  {\bibfnamefont {D.}~\bibnamefont {Branton}}, \ and\ \bibinfo {author}
  {\bibfnamefont {J.~A.}\ \bibnamefont {Golovchenko}},\ }\href {\doibase
  10.1038/nature09379} {\bibfield  {journal} {\bibinfo  {journal} {Nature}\
  }\textbf {\bibinfo {volume} {467}},\ \bibinfo {pages} {190} (\bibinfo {year}
  {2010})}\BibitemShut {NoStop}%
\bibitem [{\citenamefont {Garaj}\ \emph {et~al.}(2013)\citenamefont {Garaj},
  \citenamefont {Liu}, \citenamefont {Golovchenko},\ and\ \citenamefont
  {Branton}}]{garaj_pnas_2013}%
  \BibitemOpen
  \bibfield  {author} {\bibinfo {author} {\bibfnamefont {S.}~\bibnamefont
  {Garaj}}, \bibinfo {author} {\bibfnamefont {S.}~\bibnamefont {Liu}}, \bibinfo
  {author} {\bibfnamefont {J.~A.}\ \bibnamefont {Golovchenko}}, \ and\ \bibinfo
  {author} {\bibfnamefont {D.}~\bibnamefont {Branton}},\ }\href {\doibase
  10.1073/pnas.1220012110} {\bibfield  {journal} {\bibinfo  {journal} {Proc.
  Nat. Acad. Sci.}\ }\textbf {\bibinfo {volume} {110}},\ \bibinfo {pages}
  {12192} (\bibinfo {year} {2013})}\BibitemShut {NoStop}%
\bibitem [{\citenamefont {O'Hern}\ \emph {et~al.}(2014)\citenamefont {O'Hern},
  \citenamefont {Boutilier}, \citenamefont {Idrobo}, \citenamefont {Song},
  \citenamefont {Kong}, \citenamefont {Laoui}, \citenamefont {Atieh},\ and\
  \citenamefont {Karnik}}]{OHern_nl_2014}%
  \BibitemOpen
  \bibfield  {author} {\bibinfo {author} {\bibfnamefont {S.~C.}\ \bibnamefont
  {O'Hern}}, \bibinfo {author} {\bibfnamefont {M.~S.~H.}\ \bibnamefont
  {Boutilier}}, \bibinfo {author} {\bibfnamefont {J.-C.}\ \bibnamefont
  {Idrobo}}, \bibinfo {author} {\bibfnamefont {Y.}~\bibnamefont {Song}},
  \bibinfo {author} {\bibfnamefont {J.}~\bibnamefont {Kong}}, \bibinfo {author}
  {\bibfnamefont {T.}~\bibnamefont {Laoui}}, \bibinfo {author} {\bibfnamefont
  {M.}~\bibnamefont {Atieh}}, \ and\ \bibinfo {author} {\bibfnamefont
  {R.}~\bibnamefont {Karnik}},\ }\href {\doibase 10.1021/nl404118f} {\bibfield
  {journal} {\bibinfo  {journal} {Nano Lett.}\ }\textbf {\bibinfo {volume}
  {14}},\ \bibinfo {pages} {1234} (\bibinfo {year} {2014})}\BibitemShut
  {NoStop}%
\bibitem [{\citenamefont {Rollings}, \citenamefont {Kuan},\ and\ \citenamefont
  {Golovchenko}(2016)}]{rollings_nc_2016}%
  \BibitemOpen
  \bibfield  {author} {\bibinfo {author} {\bibfnamefont {R.~C.}\ \bibnamefont
  {Rollings}}, \bibinfo {author} {\bibfnamefont {A.~T.}\ \bibnamefont {Kuan}},
  \ and\ \bibinfo {author} {\bibfnamefont {J.~A.}\ \bibnamefont
  {Golovchenko}},\ }\href {\doibase 10.1038/ncomms11408} {\bibfield  {journal}
  {\bibinfo  {journal} {Nat. Comm.}\ }\textbf {\bibinfo {volume} {7}},\
  \bibinfo {pages} {11408} (\bibinfo {year} {2016})}\BibitemShut {NoStop}%
\bibitem [{\citenamefont {Thiruraman}\ \emph {et~al.}(2018)\citenamefont
  {Thiruraman}, \citenamefont {Fujisawa}, \citenamefont {Danda}, \citenamefont
  {Das}, \citenamefont {Zhang}, \citenamefont {Bolotsky}, \citenamefont
  {Perea-L{\'{o}}pez}, \citenamefont {Nicolaï}, \citenamefont {Senet},
  \citenamefont {Terrones},\ and\ \citenamefont
  {Drndi{\'{c}}}}]{thiruraman_nl_2018}%
  \BibitemOpen
  \bibfield  {author} {\bibinfo {author} {\bibfnamefont {J.~P.}\ \bibnamefont
  {Thiruraman}}, \bibinfo {author} {\bibfnamefont {K.}~\bibnamefont
  {Fujisawa}}, \bibinfo {author} {\bibfnamefont {G.}~\bibnamefont {Danda}},
  \bibinfo {author} {\bibfnamefont {P.~M.}\ \bibnamefont {Das}}, \bibinfo
  {author} {\bibfnamefont {T.}~\bibnamefont {Zhang}}, \bibinfo {author}
  {\bibfnamefont {A.}~\bibnamefont {Bolotsky}}, \bibinfo {author}
  {\bibfnamefont {N.}~\bibnamefont {Perea-L{\'{o}}pez}}, \bibinfo {author}
  {\bibfnamefont {A.}~\bibnamefont {Nicolaï}}, \bibinfo {author}
  {\bibfnamefont {P.}~\bibnamefont {Senet}}, \bibinfo {author} {\bibfnamefont
  {M.}~\bibnamefont {Terrones}}, \ and\ \bibinfo {author} {\bibfnamefont
  {M.}~\bibnamefont {Drndi{\'{c}}}},\ }\href {\doibase
  10.1021/acs.nanolett.7b04526} {\bibfield  {journal} {\bibinfo  {journal}
  {Nano Lett.}\ }\textbf {\bibinfo {volume} {18}},\ \bibinfo {pages} {1651}
  (\bibinfo {year} {2018})}\BibitemShut {NoStop}%
\bibitem [{\citenamefont {Thiruraman}, \citenamefont {Das},\ and\ \citenamefont
  {Drndi{\'{c}}}(2020)}]{thiruraman_acsnano_2020}%
  \BibitemOpen
  \bibfield  {author} {\bibinfo {author} {\bibfnamefont {J.~P.}\ \bibnamefont
  {Thiruraman}}, \bibinfo {author} {\bibfnamefont {P.~M.}\ \bibnamefont {Das}},
  \ and\ \bibinfo {author} {\bibfnamefont {M.}~\bibnamefont {Drndi{\'{c}}}},\
  }\href {\doibase 10.1021/acsnano.0c04716} {\bibfield  {journal} {\bibinfo
  {journal} {{ACS} Nano}\ }\textbf {\bibinfo {volume} {14}},\ \bibinfo {pages}
  {11831} (\bibinfo {year} {2020})}\BibitemShut {NoStop}%
\bibitem [{\citenamefont {M\'adai}\ \emph {et~al.}(2017)\citenamefont
  {M\'adai}, \citenamefont {Valisk\'o}, \citenamefont {Dallos},\ and\
  \citenamefont {Boda}}]{madai_jcp_2017}%
  \BibitemOpen
  \bibfield  {author} {\bibinfo {author} {\bibfnamefont {E.}~\bibnamefont
  {M\'adai}}, \bibinfo {author} {\bibfnamefont {M.}~\bibnamefont {Valisk\'o}},
  \bibinfo {author} {\bibfnamefont {A.}~\bibnamefont {Dallos}}, \ and\ \bibinfo
  {author} {\bibfnamefont {D.}~\bibnamefont {Boda}},\ }\href {\doibase
  10.1063/1.5007654} {\bibfield  {journal} {\bibinfo  {journal} {J. Chem.
  Phys.}\ }\textbf {\bibinfo {volume} {147}},\ \bibinfo {pages} {244702}
  (\bibinfo {year} {2017})}\BibitemShut {NoStop}%
\bibitem [{\citenamefont {Malasics}\ and\ \citenamefont
  {Boda}(2010)}]{malasics_jcp_2010}%
  \BibitemOpen
  \bibfield  {author} {\bibinfo {author} {\bibfnamefont {A.}~\bibnamefont
  {Malasics}}\ and\ \bibinfo {author} {\bibfnamefont {D.}~\bibnamefont
  {Boda}},\ }\href@noop {} {\bibfield  {journal} {\bibinfo  {journal} {J. Chem.
  Phys.}\ }\textbf {\bibinfo {volume} {132}},\ \bibinfo {pages} {244103}
  (\bibinfo {year} {2010})}\BibitemShut {NoStop}%
\bibitem [{\citenamefont {Boda}\ and\ \citenamefont
  {Gillespie}(2012)}]{boda_jctc_2012}%
  \BibitemOpen
  \bibfield  {author} {\bibinfo {author} {\bibfnamefont {D.}~\bibnamefont
  {Boda}}\ and\ \bibinfo {author} {\bibfnamefont {D.}~\bibnamefont
  {Gillespie}},\ }\href {\doibase 10.1021/ct2007988} {\bibfield  {journal}
  {\bibinfo  {journal} {J. Chem. Theor. Comput.}\ }\textbf {\bibinfo {volume}
  {8}},\ \bibinfo {pages} {824} (\bibinfo {year} {2012})}\BibitemShut {NoStop}%
\bibitem [{\citenamefont {Szymczyk}, \citenamefont {Zhu},\ and\ \citenamefont
  {Balannec}(2010)}]{szymczyk_jpcb_2010}%
  \BibitemOpen
  \bibfield  {author} {\bibinfo {author} {\bibfnamefont {A.}~\bibnamefont
  {Szymczyk}}, \bibinfo {author} {\bibfnamefont {H.}~\bibnamefont {Zhu}}, \
  and\ \bibinfo {author} {\bibfnamefont {B.}~\bibnamefont {Balannec}},\ }\href
  {\doibase 10.1021/jp1025575} {\bibfield  {journal} {\bibinfo  {journal} {J.
  Phys. Chem. B}\ }\textbf {\bibinfo {volume} {114}},\ \bibinfo {pages} {10143}
  (\bibinfo {year} {2010})}\BibitemShut {NoStop}%
\bibitem [{\citenamefont {Singh}\ and\ \citenamefont
  {Kumar}(2011{\natexlab{a}})}]{singh_jap_2011}%
  \BibitemOpen
  \bibfield  {author} {\bibinfo {author} {\bibfnamefont {K.~P.}\ \bibnamefont
  {Singh}}\ and\ \bibinfo {author} {\bibfnamefont {M.}~\bibnamefont {Kumar}},\
  }\href {\doibase http://dx.doi.org/10.1063/1.3656708} {\bibfield  {journal}
  {\bibinfo  {journal} {J. Appl. Phys.}\ }\textbf {\bibinfo {volume} {110}},\
  \bibinfo {eid} {084322} (\bibinfo {year} {2011}{\natexlab{a}})}\BibitemShut
  {NoStop}%
\bibitem [{\citenamefont {Singh}\ and\ \citenamefont
  {Kumar}(2011{\natexlab{b}})}]{singh_jpcb_2011}%
  \BibitemOpen
  \bibfield  {author} {\bibinfo {author} {\bibfnamefont {K.~P.}\ \bibnamefont
  {Singh}}\ and\ \bibinfo {author} {\bibfnamefont {M.}~\bibnamefont {Kumar}},\
  }\href {\doibase 10.1021/jp208309g} {\bibfield  {journal} {\bibinfo
  {journal} {J. Phys. Chem. C}\ }\textbf {\bibinfo {volume} {115}},\ \bibinfo
  {pages} {22917} (\bibinfo {year} {2011}{\natexlab{b}})}\BibitemShut {NoStop}%
\bibitem [{\citenamefont {Singh}, \citenamefont {Kumari},\ and\ \citenamefont
  {Kumar}(2011)}]{singh_apl_2011}%
  \BibitemOpen
  \bibfield  {author} {\bibinfo {author} {\bibfnamefont {K.~P.}\ \bibnamefont
  {Singh}}, \bibinfo {author} {\bibfnamefont {K.}~\bibnamefont {Kumari}}, \
  and\ \bibinfo {author} {\bibfnamefont {M.}~\bibnamefont {Kumar}},\ }\href
  {\doibase http://dx.doi.org/10.1063/1.3627181} {\bibfield  {journal}
  {\bibinfo  {journal} {Appl. Phys. Lett.}\ }\textbf {\bibinfo {volume} {99}},\
  \bibinfo {eid} {113103} (\bibinfo {year} {2011}),\
  http://dx.doi.org/10.1063/1.3627181}\BibitemShut {NoStop}%
\bibitem [{\citenamefont {van Oeffelen}\ \emph {et~al.}(2015)\citenamefont {van
  Oeffelen}, \citenamefont {Roy}, \citenamefont {Idrissi}, \citenamefont
  {Charlier}, \citenamefont {Lagae},\ and\ \citenamefont
  {Borghs}}]{oeffelen_oneplos_2015}%
  \BibitemOpen
  \bibfield  {author} {\bibinfo {author} {\bibfnamefont {L.}~\bibnamefont {van
  Oeffelen}}, \bibinfo {author} {\bibfnamefont {W.~V.}\ \bibnamefont {Roy}},
  \bibinfo {author} {\bibfnamefont {H.}~\bibnamefont {Idrissi}}, \bibinfo
  {author} {\bibfnamefont {D.}~\bibnamefont {Charlier}}, \bibinfo {author}
  {\bibfnamefont {L.}~\bibnamefont {Lagae}}, \ and\ \bibinfo {author}
  {\bibfnamefont {G.}~\bibnamefont {Borghs}},\ }\href {\doibase
  10.1371/journal.pone.0124171} {\bibfield  {journal} {\bibinfo  {journal}
  {PLoS ONE}\ }\textbf {\bibinfo {volume} {10}},\ \bibinfo {pages} {e0124171}
  (\bibinfo {year} {2015})}\BibitemShut {NoStop}%
\bibitem [{\citenamefont {Tajparast}, \citenamefont {Virdi},\ and\
  \citenamefont {Glavinovi\'{c}}(2015)}]{tajparast_bba_2015}%
  \BibitemOpen
  \bibfield  {author} {\bibinfo {author} {\bibfnamefont {M.}~\bibnamefont
  {Tajparast}}, \bibinfo {author} {\bibfnamefont {G.}~\bibnamefont {Virdi}}, \
  and\ \bibinfo {author} {\bibfnamefont {M.~I.}\ \bibnamefont
  {Glavinovi\'{c}}},\ }\href {\doibase
  http://dx.doi.org/10.1016/j.bbamem.2015.05.023} {\bibfield  {journal}
  {\bibinfo  {journal} {Biochim. Biophys. Acta (BBA) - Biomem.}\ }\textbf
  {\bibinfo {volume} {1848}},\ \bibinfo {pages} {2138} (\bibinfo {year}
  {2015})}\BibitemShut {NoStop}%
\bibitem [{\citenamefont {Gillespie}, \citenamefont {Nonner},\ and\
  \citenamefont {Eisenberg}(2002)}]{gillespie_jpcm_2002}%
  \BibitemOpen
  \bibfield  {author} {\bibinfo {author} {\bibfnamefont {D.}~\bibnamefont
  {Gillespie}}, \bibinfo {author} {\bibfnamefont {W.}~\bibnamefont {Nonner}}, \
  and\ \bibinfo {author} {\bibfnamefont {R.~S.}\ \bibnamefont {Eisenberg}},\
  }\href {\doibase 10.1088/0953-8984/14/46/317} {\bibfield  {journal} {\bibinfo
   {journal} {J. Phys. Condens. Matter}\ }\textbf {\bibinfo {volume} {14}},\
  \bibinfo {pages} {12129} (\bibinfo {year} {2002})}\BibitemShut {NoStop}%
\bibitem [{\citenamefont {Gillespie}, \citenamefont {Nonner},\ and\
  \citenamefont {Eisenberg}(2003)}]{gillespie_pre_2003}%
  \BibitemOpen
  \bibfield  {author} {\bibinfo {author} {\bibfnamefont {D.}~\bibnamefont
  {Gillespie}}, \bibinfo {author} {\bibfnamefont {W.}~\bibnamefont {Nonner}}, \
  and\ \bibinfo {author} {\bibfnamefont {R.~S.}\ \bibnamefont {Eisenberg}},\
  }\href {\doibase 10.1103/physreve.68.031503} {\bibfield  {journal} {\bibinfo
  {journal} {Phys. Rev. E}\ }\textbf {\bibinfo {volume} {68}},\ \bibinfo
  {pages} {031503} (\bibinfo {year} {2003})}\BibitemShut {NoStop}%
\bibitem [{\citenamefont {Gillespie}\ \emph {et~al.}(2005)\citenamefont
  {Gillespie}, \citenamefont {Xu}, \citenamefont {Wang},\ and\ \citenamefont
  {Meissner}}]{gillespie_jpcb_2005}%
  \BibitemOpen
  \bibfield  {author} {\bibinfo {author} {\bibfnamefont {D.}~\bibnamefont
  {Gillespie}}, \bibinfo {author} {\bibfnamefont {L.}~\bibnamefont {Xu}},
  \bibinfo {author} {\bibfnamefont {Y.}~\bibnamefont {Wang}}, \ and\ \bibinfo
  {author} {\bibfnamefont {G.}~\bibnamefont {Meissner}},\ }\href {\doibase
  10.1021/jp052471j} {\bibfield  {journal} {\bibinfo  {journal} {J. Phys. Chem.
  B}\ }\textbf {\bibinfo {volume} {109}},\ \bibinfo {pages} {15598} (\bibinfo
  {year} {2005})}\BibitemShut {NoStop}%
\bibitem [{\citenamefont {Gillespie}(2008)}]{gillespie_bj_2008_energetics}%
  \BibitemOpen
  \bibfield  {author} {\bibinfo {author} {\bibfnamefont {D.}~\bibnamefont
  {Gillespie}},\ }\href {\doibase 10.1529/biophysj.107.116798} {\bibfield
  {journal} {\bibinfo  {journal} {Biophys. J.}\ }\textbf {\bibinfo {volume}
  {94}},\ \bibinfo {pages} {1169} (\bibinfo {year} {2008})}\BibitemShut
  {NoStop}%
\bibitem [{\citenamefont {Burger}, \citenamefont {Schlake},\ and\ \citenamefont
  {Wolfram}(2012)}]{burger_nonlin_2012}%
  \BibitemOpen
  \bibfield  {author} {\bibinfo {author} {\bibfnamefont {M.}~\bibnamefont
  {Burger}}, \bibinfo {author} {\bibfnamefont {B.}~\bibnamefont {Schlake}}, \
  and\ \bibinfo {author} {\bibfnamefont {M.-T.}\ \bibnamefont {Wolfram}},\
  }\href {\doibase 10.1088/0951-7715/25/4/961} {\bibfield  {journal} {\bibinfo
  {journal} {Nonlinearity}\ }\textbf {\bibinfo {volume} {25}},\ \bibinfo
  {pages} {961} (\bibinfo {year} {2012})}\BibitemShut {NoStop}%
\bibitem [{\citenamefont {Eisenberg}(2019)}]{eisenberg_pnp_2019}%
  \BibitemOpen
  \bibfield  {author} {\bibinfo {author} {\bibfnamefont {R.}~\bibnamefont
  {Eisenberg}},\ }\href {\doibase 10.31224/osf.io/2739d} {\  (\bibinfo {year}
  {2019}),\ 10.31224/osf.io/2739d}\BibitemShut {NoStop}%
\bibitem [{\citenamefont {Boda}(2014)}]{boda_arcc_2014}%
  \BibitemOpen
  \bibfield  {author} {\bibinfo {author} {\bibfnamefont {D.}~\bibnamefont
  {Boda}},\ }in\ \href {\doibase 10.1016/b978-0-444-63378-1.00005-7} {\emph
  {\bibinfo {booktitle} {Ann. Rep. Comp. Chem.}}},\ Vol.~\bibinfo {volume}
  {10},\ \bibinfo {editor} {edited by\ \bibinfo {editor} {\bibfnamefont
  {R.~A.}\ \bibnamefont {Wheeler}}}\ (\bibinfo  {publisher} {Elsevier},\
  \bibinfo {year} {2014})\ Chap.\ \bibinfo {chapter} {5 {Monte Carlo}
  Simulation of Electrolyte Solutions in Biology: {In} and Out of Equilibrium},
  pp.\ \bibinfo {pages} {127--163}\BibitemShut {NoStop}%
\bibitem [{\citenamefont {Boda}\ \emph {et~al.}(2014)\citenamefont {Boda},
  \citenamefont {Kov\'acs}, \citenamefont {Gillespie},\ and\ \citenamefont
  {Krist\'of}}]{boda_jml_2014}%
  \BibitemOpen
  \bibfield  {author} {\bibinfo {author} {\bibfnamefont {D.}~\bibnamefont
  {Boda}}, \bibinfo {author} {\bibfnamefont {R.}~\bibnamefont {Kov\'acs}},
  \bibinfo {author} {\bibfnamefont {D.}~\bibnamefont {Gillespie}}, \ and\
  \bibinfo {author} {\bibfnamefont {T.}~\bibnamefont {Krist\'of}},\ }\href
  {\doibase 10.1016/j.molliq.2013.03.015} {\bibfield  {journal} {\bibinfo
  {journal} {J. Mol. Liq.}\ }\textbf {\bibinfo {volume} {189}},\ \bibinfo
  {pages} {100} (\bibinfo {year} {2014})}\BibitemShut {NoStop}%
\bibitem [{\citenamefont {Fertig}\ \emph {et~al.}(2017)\citenamefont {Fertig},
  \citenamefont {M\'adai}, \citenamefont {Valisk\'o},\ and\ \citenamefont
  {Boda}}]{fertig_hjic_2017}%
  \BibitemOpen
  \bibfield  {author} {\bibinfo {author} {\bibfnamefont {D.}~\bibnamefont
  {Fertig}}, \bibinfo {author} {\bibfnamefont {E.}~\bibnamefont {M\'adai}},
  \bibinfo {author} {\bibfnamefont {M.}~\bibnamefont {Valisk\'o}}, \ and\
  \bibinfo {author} {\bibfnamefont {D.}~\bibnamefont {Boda}},\ }\href {\doibase
  10.1515/hjic-2017-0011} {\bibfield  {journal} {\bibinfo  {journal} {Hung. J.
  Ind. Chem.}\ }\textbf {\bibinfo {volume} {45}},\ \bibinfo {pages} {73}
  (\bibinfo {year} {2017})}\BibitemShut {NoStop}%
\bibitem [{\citenamefont {Gillespie}\ and\ \citenamefont
  {Boda}(2008)}]{gillespie_bj_2008_ca}%
  \BibitemOpen
  \bibfield  {author} {\bibinfo {author} {\bibfnamefont {D.}~\bibnamefont
  {Gillespie}}\ and\ \bibinfo {author} {\bibfnamefont {D.}~\bibnamefont
  {Boda}},\ }\href {\doibase 10.1529/biophysj.107.127977} {\bibfield  {journal}
  {\bibinfo  {journal} {Biophys. J.}\ }\textbf {\bibinfo {volume} {95}},\
  \bibinfo {pages} {2658} (\bibinfo {year} {2008})}\BibitemShut {NoStop}%
\bibitem [{\citenamefont {Gillespie}\ \emph {et~al.}(2008)\citenamefont
  {Gillespie}, \citenamefont {Boda}, \citenamefont {He}, \citenamefont {Apel},\
  and\ \citenamefont {Siwy}}]{gillespie_bj_2008_nanopore}%
  \BibitemOpen
  \bibfield  {author} {\bibinfo {author} {\bibfnamefont {D.}~\bibnamefont
  {Gillespie}}, \bibinfo {author} {\bibfnamefont {D.}~\bibnamefont {Boda}},
  \bibinfo {author} {\bibfnamefont {Y.}~\bibnamefont {He}}, \bibinfo {author}
  {\bibfnamefont {P.}~\bibnamefont {Apel}}, \ and\ \bibinfo {author}
  {\bibfnamefont {Z.}~\bibnamefont {Siwy}},\ }\href {\doibase
  10.1529/biophysj.107.127985} {\bibfield  {journal} {\bibinfo  {journal}
  {Biophys. J.}\ }\textbf {\bibinfo {volume} {95}},\ \bibinfo {pages} {609}
  (\bibinfo {year} {2008})}\BibitemShut {NoStop}%
\bibitem [{\citenamefont {He}\ \emph {et~al.}(2009)\citenamefont {He},
  \citenamefont {Gillespie}, \citenamefont {Boda}, \citenamefont {Vlassiouk},
  \citenamefont {Eisenberg},\ and\ \citenamefont {Siwy}}]{he_jacs_2009}%
  \BibitemOpen
  \bibfield  {author} {\bibinfo {author} {\bibfnamefont {Y.}~\bibnamefont
  {He}}, \bibinfo {author} {\bibfnamefont {D.}~\bibnamefont {Gillespie}},
  \bibinfo {author} {\bibfnamefont {D.}~\bibnamefont {Boda}}, \bibinfo {author}
  {\bibfnamefont {I.}~\bibnamefont {Vlassiouk}}, \bibinfo {author}
  {\bibfnamefont {R.~S.}\ \bibnamefont {Eisenberg}}, \ and\ \bibinfo {author}
  {\bibfnamefont {Z.~S.}\ \bibnamefont {Siwy}},\ }\href {\doibase
  10.1021/ja808717u} {\bibfield  {journal} {\bibinfo  {journal} {JACS}\
  }\textbf {\bibinfo {volume} {131}},\ \bibinfo {pages} {5194} (\bibinfo {year}
  {2009})}\BibitemShut {NoStop}%
\bibitem [{\citenamefont {Boda}\ \emph {et~al.}(2009)\citenamefont {Boda},
  \citenamefont {Valisk{\'o}}, \citenamefont {Henderson}, \citenamefont
  {Eisenberg}, \citenamefont {Gillespie},\ and\ \citenamefont
  {Nonner}}]{boda_jgp_2009}%
  \BibitemOpen
  \bibfield  {author} {\bibinfo {author} {\bibfnamefont {D.}~\bibnamefont
  {Boda}}, \bibinfo {author} {\bibfnamefont {M.}~\bibnamefont {Valisk{\'o}}},
  \bibinfo {author} {\bibfnamefont {D.}~\bibnamefont {Henderson}}, \bibinfo
  {author} {\bibfnamefont {B.}~\bibnamefont {Eisenberg}}, \bibinfo {author}
  {\bibfnamefont {D.}~\bibnamefont {Gillespie}}, \ and\ \bibinfo {author}
  {\bibfnamefont {W.}~\bibnamefont {Nonner}},\ }\href {\doibase
  10.1085/jgp.200910211} {\bibfield  {journal} {\bibinfo  {journal} {J. Gen.
  Physiol.}\ }\textbf {\bibinfo {volume} {133}},\ \bibinfo {pages} {497}
  (\bibinfo {year} {2009})}\BibitemShut {NoStop}%
\bibitem [{\citenamefont {Malasics}\ \emph {et~al.}(2010)\citenamefont
  {Malasics}, \citenamefont {Boda}, \citenamefont {Valisk{\'o}}, \citenamefont
  {Henderson},\ and\ \citenamefont {Gillespie}}]{malasics_bba_2010_trivalent}%
  \BibitemOpen
  \bibfield  {author} {\bibinfo {author} {\bibfnamefont {M.}~\bibnamefont
  {Malasics}}, \bibinfo {author} {\bibfnamefont {D.}~\bibnamefont {Boda}},
  \bibinfo {author} {\bibfnamefont {M.}~\bibnamefont {Valisk{\'o}}}, \bibinfo
  {author} {\bibfnamefont {D.}~\bibnamefont {Henderson}}, \ and\ \bibinfo
  {author} {\bibfnamefont {D.}~\bibnamefont {Gillespie}},\ }\href {\doibase
  10.1016/j.bbamem.2010.08.001} {\bibfield  {journal} {\bibinfo  {journal}
  {Biochim. et Biophys. Acta - Biomembranes}\ }\textbf {\bibinfo {volume}
  {1798}},\ \bibinfo {pages} {2013} (\bibinfo {year} {2010})}\BibitemShut
  {NoStop}%
\bibitem [{\citenamefont {Valisk\'o}\ \emph {et~al.}(2019)\citenamefont
  {Valisk\'o}, \citenamefont {Matejczyk}, \citenamefont {Hat\'o}, \citenamefont
  {Krist\'of}, \citenamefont {M\'adai}, \citenamefont {Fertig}, \citenamefont
  {Gillespie},\ and\ \citenamefont {Boda}}]{valisko_jcp_2019}%
  \BibitemOpen
  \bibfield  {author} {\bibinfo {author} {\bibfnamefont {M.}~\bibnamefont
  {Valisk\'o}}, \bibinfo {author} {\bibfnamefont {B.}~\bibnamefont
  {Matejczyk}}, \bibinfo {author} {\bibfnamefont {Z.}~\bibnamefont {Hat\'o}},
  \bibinfo {author} {\bibfnamefont {T.}~\bibnamefont {Krist\'of}}, \bibinfo
  {author} {\bibfnamefont {E.}~\bibnamefont {M\'adai}}, \bibinfo {author}
  {\bibfnamefont {D.}~\bibnamefont {Fertig}}, \bibinfo {author} {\bibfnamefont
  {D.}~\bibnamefont {Gillespie}}, \ and\ \bibinfo {author} {\bibfnamefont
  {D.}~\bibnamefont {Boda}},\ }\href {\doibase 10.1063/1.5091789} {\bibfield
  {journal} {\bibinfo  {journal} {J. Chem. Phys.}\ }\textbf {\bibinfo {volume}
  {150}},\ \bibinfo {pages} {144703} (\bibinfo {year} {2019})}\BibitemShut
  {NoStop}%
\bibitem [{\citenamefont {Boda}, \citenamefont {Valisk{\'{o}}},\ and\
  \citenamefont {Gillespie}(2020)}]{boda_entropy_2020}%
  \BibitemOpen
  \bibfield  {author} {\bibinfo {author} {\bibfnamefont {D.}~\bibnamefont
  {Boda}}, \bibinfo {author} {\bibfnamefont {M.}~\bibnamefont {Valisk{\'{o}}}},
  \ and\ \bibinfo {author} {\bibfnamefont {D.}~\bibnamefont {Gillespie}},\
  }\href {\doibase 10.3390/e22111259} {\bibfield  {journal} {\bibinfo
  {journal} {Entropy}\ }\textbf {\bibinfo {volume} {22}},\ \bibinfo {pages}
  {1259} (\bibinfo {year} {2020})}\BibitemShut {NoStop}%
\bibitem [{\citenamefont {Levine}\ \emph {et~al.}(1975)\citenamefont {Levine},
  \citenamefont {Marriott}, \citenamefont {Neale},\ and\ \citenamefont
  {Epstein}}]{levine_jcis_1975}%
  \BibitemOpen
  \bibfield  {author} {\bibinfo {author} {\bibfnamefont {S.}~\bibnamefont
  {Levine}}, \bibinfo {author} {\bibfnamefont {J.~R.}\ \bibnamefont
  {Marriott}}, \bibinfo {author} {\bibfnamefont {G.}~\bibnamefont {Neale}}, \
  and\ \bibinfo {author} {\bibfnamefont {N.}~\bibnamefont {Epstein}},\ }\href
  {\doibase 10.1016/0021-9797(75)90310-0} {\bibfield  {journal} {\bibinfo
  {journal} {J. Coll. Interf. Sci.}\ }\textbf {\bibinfo {volume} {52}},\
  \bibinfo {pages} {136} (\bibinfo {year} {1975})}\BibitemShut {NoStop}%
\bibitem [{\citenamefont {Balme}\ \emph {et~al.}(2015)\citenamefont {Balme},
  \citenamefont {Picaud}, \citenamefont {Manghi}, \citenamefont {Palmeri},
  \citenamefont {Bechelany}, \citenamefont {Cabello-Aguilar}, \citenamefont
  {Abou-Chaaya}, \citenamefont {Miele}, \citenamefont {Balanzat},\ and\
  \citenamefont {Janot}}]{balme_sr_2015}%
  \BibitemOpen
  \bibfield  {author} {\bibinfo {author} {\bibfnamefont {S.}~\bibnamefont
  {Balme}}, \bibinfo {author} {\bibfnamefont {F.}~\bibnamefont {Picaud}},
  \bibinfo {author} {\bibfnamefont {M.}~\bibnamefont {Manghi}}, \bibinfo
  {author} {\bibfnamefont {J.}~\bibnamefont {Palmeri}}, \bibinfo {author}
  {\bibfnamefont {M.}~\bibnamefont {Bechelany}}, \bibinfo {author}
  {\bibfnamefont {S.}~\bibnamefont {Cabello-Aguilar}}, \bibinfo {author}
  {\bibfnamefont {A.}~\bibnamefont {Abou-Chaaya}}, \bibinfo {author}
  {\bibfnamefont {P.}~\bibnamefont {Miele}}, \bibinfo {author} {\bibfnamefont
  {E.}~\bibnamefont {Balanzat}}, \ and\ \bibinfo {author} {\bibfnamefont
  {J.~M.}\ \bibnamefont {Janot}},\ }\href {\doibase 10.1038/srep10135}
  {\bibfield  {journal} {\bibinfo  {journal} {Sci. Rep.}\ }\textbf {\bibinfo
  {volume} {5}},\ \bibinfo {pages} {10135} (\bibinfo {year}
  {2015})}\BibitemShut {NoStop}%
\bibitem [{\citenamefont {Uematsu}\ \emph {et~al.}(2018)\citenamefont
  {Uematsu}, \citenamefont {Netz}, \citenamefont {Bocquet},\ and\ \citenamefont
  {Bonthuis}}]{uematsu_jpcb_2018}%
  \BibitemOpen
  \bibfield  {author} {\bibinfo {author} {\bibfnamefont {Y.}~\bibnamefont
  {Uematsu}}, \bibinfo {author} {\bibfnamefont {R.~R.}\ \bibnamefont {Netz}},
  \bibinfo {author} {\bibfnamefont {L.}~\bibnamefont {Bocquet}}, \ and\
  \bibinfo {author} {\bibfnamefont {D.~J.}\ \bibnamefont {Bonthuis}},\ }\href
  {\doibase 10.1021/acs.jpcb.8b01975} {\bibfield  {journal} {\bibinfo
  {journal} {J. Phys. Chem. B}\ }\textbf {\bibinfo {volume} {122}},\ \bibinfo
  {pages} {2992} (\bibinfo {year} {2018})}\BibitemShut {NoStop}%
\bibitem [{\citenamefont {Green}(2021)}]{green_jcp_2021}%
  \BibitemOpen
  \bibfield  {author} {\bibinfo {author} {\bibfnamefont {Y.}~\bibnamefont
  {Green}},\ }\href {\doibase 10.1063/5.0037873} {\bibfield  {journal}
  {\bibinfo  {journal} {J. Chem. Phys.}\ }\textbf {\bibinfo {volume} {154}},\
  \bibinfo {pages} {084705} (\bibinfo {year} {2021})}\BibitemShut {NoStop}%
\end{thebibliography}%

\end{document}